\documentclass[useAMS,usenatbib]{mnras}  
\usepackage{graphics}
\usepackage{times}
\usepackage{amsmath,amssymb}
\usepackage{bm}
\usepackage{color}
\usepackage{comment}

\def\be{\begin{equation}}
\def\ee{\end{equation}}
\def\ba{\begin{eqnarray}}
\def\ea{\end{eqnarray}}

\def\lsim{\mathrel{\rlap{\lower4pt\hbox{\hskip1pt$\sim$}}
    \raise1pt\hbox{$<$}}}                
\def\gsim{\mathrel{\rlap{\lower4pt\hbox{\hskip1pt$\sim$}}
    \raise1pt\hbox{$>$}}}

\pagerange{\pageref{firstpage}--\pageref{lastpage}}
\pubyear{2014}

\begin{document}

\label{firstpage}

\title[Cosmology using angle-only cosmic shear]{Constraining cosmology using galaxy position angle-only cosmic shear}
\author[Lee Whittaker]{Lee Whittaker\thanks{E-mail: lee.whittaker@ucl.ac.uk}\\ 
  Department of Physics and Astronomy, University College London, Gower Street, London WC1E 6BT, UK\\
  Jodrell Bank Centre for Astrophysics, School of Physics and Astronomy, University of Manchester, Oxford Road, Manchester M13 9PL, UK} \date{\today}

\maketitle

\begin{abstract}
We investigate cosmological parameter inference from realistic simulated weak lensing image data using only galaxy position angles, as opposed to full-ellipticity information. We demonstrate that input shear fields can be accurately reconstructed using only the statistics of source galaxy position angles and that, from these shear fields, we can successfully recover power spectra and infer the input cosmology. This paper builds on previous work on angle-only weak lensing estimation by extending the method to deal with variable and anisotropic PSF convolution and variable shear fields. Previous work employed a weighting scheme to downweight the contribution to shear estimates from sources aligned with the PSF. This work removes the need to downweight sources by convolving them with an image of the PSF rotated by $90^{\circ}$. We show thats this successfully undoes the rotation caused by PSF convolution, assuming we have reliable images of the PSF. We find that we can accurately recover the input shear signal from a simulated Stage III-like weak lensing data set using only the position angles to within an overall scale factor, and that the scale factor can be determined using a cosmology independent simulation with noise, galaxy, and PSF properties that match the observed data set. We then demonstrate that we can constrain cosmological parameters using angle-only shear estimates with a constraining power comparable to using current state-of-the-art shape measurement techniques that provide full-ellipticity information.
\end{abstract}

\section{Introduction}
\label{sec:intro}
Cosmic shear is the weak gravitational lensing of background source galaxies by the foreground large-scale structure of the Universe. The effects of cosmic shear can be observed as correlations in source galaxy shapes which can be used to measure the statistics of the foreground matter distribution. Moreover, by studying the distribution of matter at various redshifts, we can gain an insight into the expansion history of the Universe and explore the nature of dark energy (see \cite{bartelmann01} \citealt{bartelmann01,kilbinger15,mandelbaum18} for comprehensive reviews of weak lensing and cosmic shear). Constraining the dark energy equation of state is among the primary goals of many current cosmological missions, such as the Kilo-Degree-Survey (KiDS)\footnote{\url{http://kids.strw.leidenuniv.nl}}, the Dark Energy Survey (DES)\footnote{\url{https://www.darkenergysurvey.org}} and the Hyper Suprime-Cam Subaru Strategic Program\footnote{\url{https://hsc.mtk.nao.ac.jp/ssp/}}, and many future projects, such as the Euclid Space Mission\footnote{\url{https://www.euclid-ec.org}}, the Large Synoptic Survey Telescope (LSST)\footnote{\url{https://www.lsst.org}}, the Wide Field Infrared Survey Telescope (WFIRST) \footnote{\url{https://wfirst.gsfc.nasa.gov}} and the Chinese Space Station Optical Survey (CSS-OS) \citep{gong19}.

To date, the most common method for extracting cosmological information from weak lensing surveys involves accurately measuring the shapes of source galaxies. The shape of a galaxy can be expressed as an ellipticity which provides an estimate of the local weak lensing shear signal. Correlations in the shear as a function of spatial separation can then be quantified using a two-point correlation function (2PCF) or a power spectrum. From these two-point statistics, constraints on cosmology can be made using Bayesian inference.

There are primarily two approaches to measuring the shape of a galaxy: inverse methods, whereby properties of the galaxy are measured and then corrections applied for observational systematics, and forward model based methods that fit a parametrised model of the source to the image data. Both of these approaches are susceptible to systematics that affect both the galaxy's shape and its orientation. One of the most problematic sources of systematic is PSF convolution. This is particularly true for inverse methods based on quadrupole moments, such as KSB (\citealt{kaiser95, viola11}) and DIEMOS \citep{melchior11}, which require specific assumptions to be made about the PSF model in order to account for its effect when estimating the shear. For the simple case of an axially symmetric PSF, the galaxy is isotropically smeared, rendering an elliptical object more circular. However, when one increases the complexity to that of an elliptical PSF, both the shape and orientation of the galaxy are changed.

Model based methods, such as {\tt LENSFIT} \citealt{miller07,kitching08,miller13} and {\tt IM3SHAPE}\footnote{\url{https://bitbucket.org/joezuntz/im3shape-git}} \citep{zuntz13}, make assumptions about the models fitted to galaxy images. At the most conservative level, the assumption that galaxies are elliptical in shape is usually adopted. This is also a potential source of systematic since, in reality, galaxies have more complex morphologies. In the Third Gravitational Lensing Accuracy Testing (GREAT3) challenge\footnote{\url{http://great3challenge.info}}, this source of bias was shown to be at a level of $\sim$1\% \citep{mandelbaum15}, which is the target level for current weak lensing missions.

Noise biases affect both forward modelling and inverse methods. These biases are usually the most significant and are typically $\sim$10\% if not corrected for (\citealt{hirata03,Kacprzak12}). For moments based inverse methods, noise biases arise because ellipticity is a nonlinear function of the noisy intensity profile \citep{melchior12}. In order to mitigate the effects of noise, a weight function can be applied to the galaxy image before calculating the quadrupole moments. One must then account for the weight function to recover an estimate of the unweighted source shape (\citealt{kaiser95, viola11}). For model based methods, noise can skew the surface of the likelihood which biases inferred shape parameters. This can be addressed if one can correctly account for the effects of noise in the posterior, which is the approach adopted by Bayesian shape measurement methods, such as {\tt LENSFIT}.

In recent surveys, a variety of model and moments based shape measurement methods have been used. Galaxy shapes in the Canada-France-Hawaii-Telescope Lensing Survey (CFHTLenS) \citep{heymans12} were measured using the Bayesian algorithm {\tt LENSFIT}. Correlations in galaxy shapes, quantified using 2PCFs, were then used to constrain various cosmological parameters, both on there own and combined with external Cosmic Microwave Background (CMB) and Baryonic Acoustic Oscillation (BAO) information (\citealt{kilbinger13,heymans13,joudaki16}). Using $450\,\mathrm{deg}^2$ of KiDS data and an improved \emph{self-calibrating} version of {\tt LENSFIT} \citep{conti17}, \cite{hildebrandt17}  and \cite{kohlinger17} respectively placed constraints on cosmology using 2PCFs and a quadratic power spectrum estimator. The Dark Energy Survey Year one (DES Y1) cosmic shear analysis used 2PCFs \citep{troxel17} with two methods for estimating the shapes of galaxies \citep{zuntz18}. The primary method used \emph{metacalibration} (\citealt{huff17, sheldon17}), whereby a galaxy image is sheared by a small, known value to gauge the response of a shape estimator to the shear. The shape estimator used for this analysis was {\tt NGMIX}\footnote{\url{https://github.com/esheldon/ngmix}}, which is a model based method that fits mixtures of Gaussian profiles to a galaxy image. The second method used for DES Y1 was {\tt IM3SHAPE}. For their weak lensing analysis of the HSC first year data, \cite{mandelbaum18a} used \emph{re-Gaussianization} \citep{hirata03} to correct measured moments for the effects of the PSF. From these shapes, constraints on cosmology were made by measuring shear power spectra using a flat-sky pseudo-$C_l$ estimator \citep{hikage18} and, following this, 2PCFs \citep{hamana19}.

Measuring the shapes of galaxies to the precisions required for future Stage IV weak lensing missions, such as with Euclid and the LSST, is the target for many people working on weak lensing shape measurement, and there are many algorithms currently being developed to achieve this \citep{mandelbaum15}. It is useful to have more than one method available for estimating the shear so that systematics, which should be different for the two methods, may be better understood if the results are inconsistent. Systematics that affect angle-only estimators will not depend on ellipticity modulus information, and therefore this approach may also be complimentary to shear estimators that use full-ellipticity information when searching for systematics. 

\cite{whittaker14} developed an approach to weak lensing that focuses on using only the statistics of galaxy orientations to estimate the shear assuming an accurate knowledge of the intrinsic ellipticity distribution. This method was originally suggested as a possible way to avoid multiplicative biases in shear estimates and was successfully demonstrated to work on simple galaxy images in the absence of PSF convolution. 

The method was further developed in \cite{whittaker15a}, where it was applied to the control-ground-constant (CGC) branch of the GREAT3 simulations \citep{mandelbaum14}. These simulations consist of 200 fields, and each field contains 10,000 sources and a constant shear signal. Each source within a field is convolved with the same anisotropic PSF, and high signal-to-noise images of the PSF are provided. Three methods were used to estimate the position angles: the integrated light method (see \cite{whittaker15a} for details), measuring quadrupole moments, and model fitting using {\tt IM3SHAPE}. The PSF images were included in the fitted model when using {\tt IM3SHAPE}. The integrated light and moments based methods are model independent. To account for PSF anisotropy when using these two techniques, a weighting scheme was adopted, whereby the contribution to shear estimates from galaxies aligned with the PSF was downweighted. This weighting scheme requires many simulated images to accurately construct a weighting function for each PSF model. There is also expected to be an increase in  errors on the shear estimates, as downweighting reduces information.

For each position angle measurement method applied to the GREAT3 simulations, the angle-only shear estimator was shown to perform better than the highest entry to the GREAT3 challenge using KSB and comparable to the highest entry using {\tt IM3SHAPE} with full-ellipticity information.

In this paper, we mitigate the effects of PSF anisotropy by convolving each galaxy image with an image of the PSF rotated by $90^{\circ}$. Assuming we have an accurate image of the PSF at the position of the source, this convolution results in an isotropically smeared galaxy image, and so it undoes the unwanted rotational effects of the PSF. This approach does not require us to downweight contributions from sources aligned with the PSF, and so we are not required to construct weight functions using simulations. There is a small increase in the errors on the position angle estimates, since the ellipticities of the galaxies are slightly reduced by the extra convolution; however, this increase is found to be small in our simulations. Since we are not concerned with the moduli of galaxy ellipticities for angle-only shear estimation, the circularisation of the source is not a problem.

The main goal of this paper is to show that cosmological parameters can be competitively inferred from weak lensing surveys using only galaxy position angle measurements. To do this, we simulate fields of Stage III-like galaxy images with an input shear signal, realistic noise levels, and a variable anisotropic PSF. From the simulated images, we measure the position angles of the galaxies using the rotated PSF convolution method discussed above. These position angles are used to estimate the shear signal, from which the input cosmology is inferred. As a comparison, we also use {\tt IM3SHAPE} to estimate shapes from the images, and similar analyses are carried out using both full-ellipticity information and only position angles calculated from the {\tt IM3SHAPE} ellipticities.

We begin by briefly reviewing previous work on angle-only shear estimation in Section \ref{ao:shear}. The method for estimating a galaxy’s position angle using a $90^{\circ}$ rotated image of the PSF is discussed in Section \ref{sec:meas}. In Section \ref{sec:cos_sim}, we describe the Stage III weak lensing simulation. In Section \ref{sec:est_shear}, we discuss estimating shears from the simulation and look at the performance of the angle-only estimator. The shear power spectra are presented in Section \ref{sec:ps}. The likelihood and covariance matrix used to infer cosmology are discussed in Section \ref{sec:likelihood}. The main results of the paper are in Section \ref{sec:results}. Finally, we conclude in Section \ref{sec:conc}.

\section{Angle-only weak lensing}
\label{ao:shear}
The approach to estimating a weak lensing shear signal using only galaxy orientations was first proposed by \cite{kochanek90}, who suggested that such estimates should be free from the systematic effects of isotropic smearing by an atmospheric PSF. This idea was explored in more detail in \cite{schneider95}, where it was shown that information about the modulus of the shear can be extracted from the statistics of measured position angles if one has
a prior knowledge of the intrinsic ellipticity distribution. The approach was developed further in \cite{whittaker14}, where an estimator for the shear was proposed and demonstrated on various simulations, and on the already calibrated galaxy shapes in the publicly available CHTLenS shear catalogue\footnote{\url{http://www.cfhtlens.org}} \citep{miller13}.

Here, we briefly summarise the formalism of angle-only shear estimation presented in \cite{whittaker14} and describe the approach adopted for estimating shear from the Stage III simulation discussed later in this paper.

If we define the second-order moments of a galaxy's intensity profile, $I_{\mathrm{gal}}(\bm{\theta})$, as
\begin{equation}
Q_{MN}=\int\mathrm{d}^2\bm{\theta}\left(x-\bar{x}\right)^M\left(y-\bar{y}\right)^NI_{\mathrm{gal}}\left(\bm{\theta}\right),
\end{equation}
where $\bm{\theta}=(x,y)$ denotes position on the sky and the bar indicates the coordinates of the centroid of $I_{\mathrm{gal}}(\bm{\theta})$, then the complex ellipticity ($\epsilon\equiv\epsilon_1+i\epsilon_2$) of the galaxy can be defined as
\begin{equation}\label{eq:def_ellip}
\epsilon=\frac{Q_{20}-Q_{02}+i2Q_{11}}{Q_{20}+Q_{02}+2\sqrt{Q_{20}Q_{02}-Q_{11}^2}}.
\end{equation}
This definition of ellipticity is one of two definitions commonly used in weak lensing, the other being
\begin{equation}\label{eq:chi_ellip}
\chi=\frac{Q_{20}-Q_{02}+i2Q_{11}}{Q_{20}+Q_{02}}.
\end{equation}
The position angle of a galaxy, $\alpha$, is identical for both of these definitions, and is given by
\begin{equation}\label{eq:pos_angle}
\alpha=\frac{1}{2}\tan^{-1}\left(\frac{2Q_{11}}{Q_{20}-Q_{02}}\right).
\end{equation}

Assuming the definition of ellipticity given in equation (\ref{eq:def_ellip}), the lensed ellipticity of a source, $\epsilon^{\mathrm{lens}}$, can be written in terms of the intrinsic ellipticity, $\epsilon^{\mathrm{int}}$, and the reduced shear, $g$, such that
\begin{equation}\label{eq:lensed_ellip}
\epsilon^{\mathrm{lens}}=\frac{\epsilon^{\mathrm{int}}+g}{1+g^*\epsilon^{\mathrm{int}}},
\end{equation}
where the asterisk denotes complex conjugation. The observed ellipticity, $\epsilon^{\mathrm{obs}}$, will also have a contribution from measurement errors, $\delta_{\epsilon}$:
\begin{equation}
\epsilon^{\mathrm{obs}}=\epsilon^{\mathrm{lens}}+\delta_{\epsilon}.
\end{equation}

If we assume that galaxies are intrinsically randomly orientated ($<\epsilon^{\mathrm{int}}>=0$) and that we have unbiased measurements of galaxy ellipticities, an unbiased estimator for the shear can be written as
\begin{equation}\label{eq:st_est}
\hat{g}=\frac{\sum_{i=1}^Nw_i\epsilon_{(i)}^{\mathrm{obs}}}{\sum_{i=1}^Nw_i},
\end{equation}
where the summations are over all galaxies in a region of sky small enough that the shear can be considered constant. This is the standard method for estimating the shear using full-ellipticity information.

The position angles of the galaxies can estimated either directly from the image data using, for example, quadrupole moments (equation (\ref{eq:pos_angle})) or indirectly using a set of ellipticity measurements.

We can write the trigonometric functions of estimated position angles, $\hat{\alpha}$, in terms of the shear position angle, $\alpha_0$, and an offset, $\delta$, where
\begin{align}\label{eq:a0_delta}
\cos\left(2n\hat{\alpha}\right)=&\cos\left(2n\alpha_0+2n\delta\right),\nonumber\\
\sin\left(2n\hat{\alpha}\right)=&\sin\left(2n\alpha_0+2n\delta\right),
\end{align}
and where $n$ is an integer. The angle $\delta$ will have contributions from both the intrinsic galaxy shape and measurement errors. 

If one assumes that $<\epsilon^{\mathrm{int}}>=0$ and that measurement errors on the angles are symmetric about zero, the distribution of $\delta$ will be symmetric about zero, but the shape and dispersion of the distribution will depend on the modulus of the shear, the error distribution, and the intrinsic ellipticity distribution. In this case, the expectation values of the trigonometric functions are
\begin{align}
\left<\cos\left(2n\hat{\alpha}\right)\right>=&\left<\cos\left(2n\alpha_0+2n\delta\right)\right>,\nonumber\\
=&\left<\cos\left(2n\delta\right)\right>\cos\left(2n\alpha_0\right),\nonumber\\
\left<\sin\left(2n\hat{\alpha}\right)\right>=&\left<\sin\left(2n\alpha_0+2n\delta\right)\right>,\nonumber\\
=&\left<\cos\left(2n\delta\right)\right>\sin\left(2n\alpha_0\right),
\end{align}
where $\left<\cos\left(2n\delta\right)\right>$ is a function of the shear,
\begin{equation}\label{eq:Fn_cosdelta}
F_n\left(\left| g\right|\right)\equiv\left<\cos\left(2n\delta\right)\right>.
\end{equation}

In \cite{whittaker14} and \cite{whittaker15a}, the $F_1$ function included information about the intrinsic ellipticity distribution only, in which case errors on the position angle measurements were found to produce biases in the shear estimates which were then calibrated for. In this work, errors are included in the distribution of $\delta$, and so in the $F_1$ function. An expression for $F_1$ for the case of zero measurement errors was presented in \cite{whittaker14} and is given in Appendix \ref{ap:F1}, where it is also extended to include measurement errors.

Given a set of galaxy position angles, $\hat{\alpha}^{(i)}$, in a region of sky small enough that we can assume a constant shear signal, the angle-only shear estimator proposed by \cite{whittaker14} is
\begin{align}\label{eq:full_est}
\hat{\alpha}_0=&\frac{1}{2}\tan^{-1}\left[\frac{\sum_{i=1}^N\sin\left(2\hat{\alpha}^{(i)}\right)}{\sum_{i=1}^N\cos\left(2\hat{\alpha}^{(i)}\right)}\right],\nonumber\\
F_1\left(\left|\hat{g}\right|\right)=&\sqrt{\left[\frac{1}{N}\sum_{i=1}^N\cos\left(2\hat{\alpha}^{(i)}\right)\right]^2 + \left[\frac{1}{N}\sum_{i=1}^N\sin\left(2\hat{\alpha}^{(i)}\right)\right]^2},
\end{align}
where the summations are over all $N$ galaxies in the small region of sky and an estimate of the modulus of the shear can be recovered by inverting the function $F_1\left(\left|\hat{g}\right|\right)$. Throughout this paper hats are used to denote estimated quantities.

If $<\epsilon^{\mathrm{int}}>=0$ and measurement errors on the position angles are symmetric about zero, the $F_n$ functions must be zero for the case of zero shear, since the distribution of $\delta$ in equation (\ref{eq:Fn_cosdelta}) will be uniform. In this case, if $|g|$ is much less than the dispersion in galaxy ellipticities, the $F_1$ function can be expanded to first order in $|g|$ as
\begin{equation}\label{eq:approx_F}
F_1\left(\left|g\right|\right)\approx\frac{\left|g\right|}{\mu},
\end{equation}
where $\mu$ is a constant that depends on the distribution of $\epsilon^{\mathrm{int}}$ and the distribution of errors on the position angle measurements. The estimator in equation (\ref{eq:full_est}) can then be simplified as
\begin{align}\label{eq:linest}
\hat{g}_1=&\frac{\mu}{N}\sum_{i=1}^N\cos\left(2\hat{\alpha}^{(i)}\right),\nonumber\\
\hat{g}_2=&\frac{\mu}{N}\sum_{i=1}^N\sin\left(2\hat{\alpha}^{(i)}\right).
\end{align}
These are the forms of the $F_1$ function and shear estimator assumed for this paper. For our analysis of the Stage III simulation, we estimate $\mu$ using a simulated calibration data set containing randomly orientated input shears with constant $|g|$; this is discussed in Section \ref{subsec:rec_shear}.

\begin{figure*}
\begin{minipage}{6in}
\centering
\includegraphics{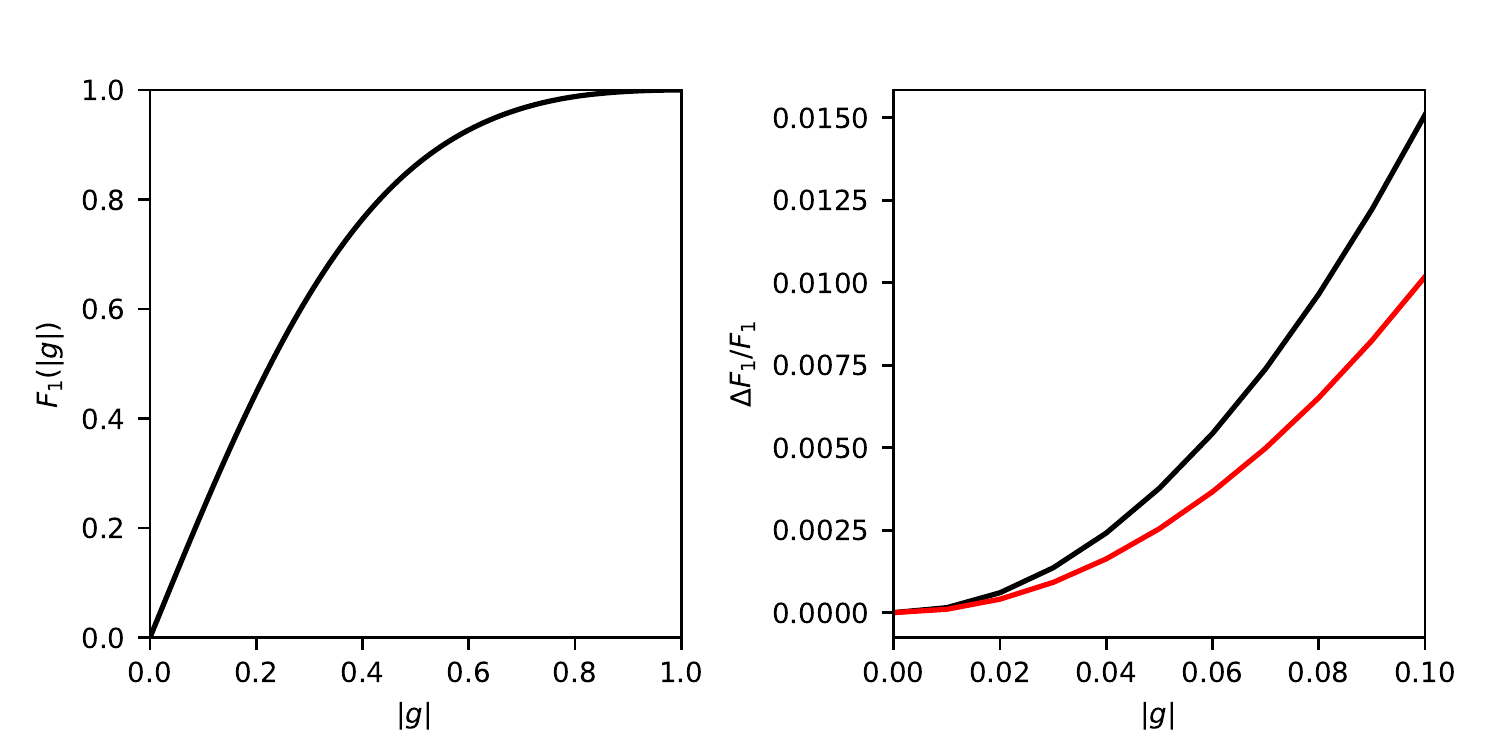}
\caption{\emph{Left panel}: An example of the $F_1$ function (equation (\ref{eq:general_F}))) for Rayleigh distributed $|\epsilon^{\mathrm{int}}|$ (equation (\ref{eq:rayleigh_e})) and assuming zero measurement errors on the position angles. In the \emph{right panel}, the black curve shows the fractional difference between the $F_1$ function and the first-order approximation (equation (\ref{eq:approx_F})). We see that the difference between the two functions increases with $|g|$ and reaches 1\% at $g\sim$0.08. The red curve shows the case where we increase $\sigma_{\epsilon}$ from 0.29 to 0.36 to model for measurement errors. We see that the accuracy of the linear approximation has increased with increased $\sigma_{\epsilon}$.}
\label{fig:deltaF}
\end{minipage}
\end{figure*}

An example of the full $F_1$ function is given in the left-hand panel of Fig. \ref{fig:deltaF} for the case where we assume zero measurement errors and that $|\epsilon^{\mathrm{int}}|$ is Rayleigh distributed with maximum ellipticity
$|\epsilon^{\mathrm{int}}_{\mathrm{max}}|=0.804$ and dispersion parameter $\sigma_{\mathrm{\epsilon}}=0.29$. In this case, the normalised intrinsic ellipticity distribution is
\begin{equation}\label{eq:rayleigh_e}
f\left(\left|\epsilon^{\mathrm{int}}\right|\right)=\frac{1}{\left[1-\exp\left(-\frac{\left|\epsilon^{\mathrm{int}}_{\mathrm{max}}\right|^2}{2\sigma_{\epsilon}^2}\right)\right]\sigma_{\epsilon}^2}\exp\left(-\frac{\left|\epsilon^{\mathrm{int}}\right|^2}{2\sigma_{\epsilon}^2}\right).
\end{equation}
This is the intrinsic ellipticity distribution assumed for the simulated data set discussed in Section \ref{sec:cos_sim}.

In the right-hand panel of Fig. \ref{fig:deltaF} , we show the fractional difference between the full $F_1$ function (given in equation (\ref{eq:general_F})) and the first-order approximation, given in equation (\ref{eq:approx_F}). Here $\Delta F_1=|g|/\mu-F_1$. The value of $\mu$ is calculated by Taylor expanding the full $F_1$ function, and an expression for $\mu$ is given in equation (\ref{eq:taylor_k}). It should be emphasised that this $F_1$ function assumes zero measurement errors, whereas for the simulation in Section \ref{sec:cos_sim}, measurement errors are included. In order to look at the impact of measurement errors on the $F_1$ function, we approximate the effects by increasing $\sigma_{\mathrm{\epsilon}}$ to 0.36\footnote{This approximation assumes that $\epsilon^{\mathrm{obs}}\approx\epsilon^{\mathrm{int}}+g+\delta_{\epsilon}.$} and again look at $\Delta F1/F1$. This is shown as the red curve in the right-hand panel of Fig. \ref{fig:deltaF}. In this case, we see that the accuracy of the first-order approximation has increased. This increase in accuracy is likely due to ellipticities being much larger than the shear for an increased number of sources when the dispersion is increased.

The accuracy with which equation (\ref{eq:approx_F}) describes the full $F_1$ function depends on the form of $F_1$ and so on the distributions of intrinsic ellipticities and measurement errors. In a more general scenario, it should be possible to construct the full $F_1$ function using calibration simulations. This is discussed in more detail in Section \ref{subsec:rec_shear}.

\section{Measuring the position angles}
\label{sec:meas}
In this section, we develop a model independent method for measuring position angles of sources which, in principle, undoes the unwanted rotation effects of anisotropic PSF convolution by further convolving the observed galaxy image with an image of the PSF rotated by $90^{\circ}$. Throughout this paper, we will refer to this method for undoing PSF rotation as the Rotated PSF Convolution (RPC) method. By removing the effects of PSF anisotropy from the position angles, we avoid having to downweight the contribution from sources aligned with the PSF when estimating the shear using the angle-only estimator.

In Appendix \ref{ap:rot_convol}, we show analytically that such a convolution should undo the rotation caused by PSF anisotropy for the ideal case of zero noise and when we ignore pixelation. This is a general result that holds for any galaxy and PSF morphology, provided that their quadrupole moments do not diverge, which is the case for all non-pathological distributions \citep{melchior11}. The extra convolution results in an isotropically smeared image of the source galaxy, and so ellipticity measurements, which are defined as a nonlinear combination of quadrupole moments (see equation (\ref{eq:def_ellip})), made using the resulting image would still be biased if they did not account for this effect.

\begin{figure*}
\begin{minipage}{6in}
\centering
\includegraphics{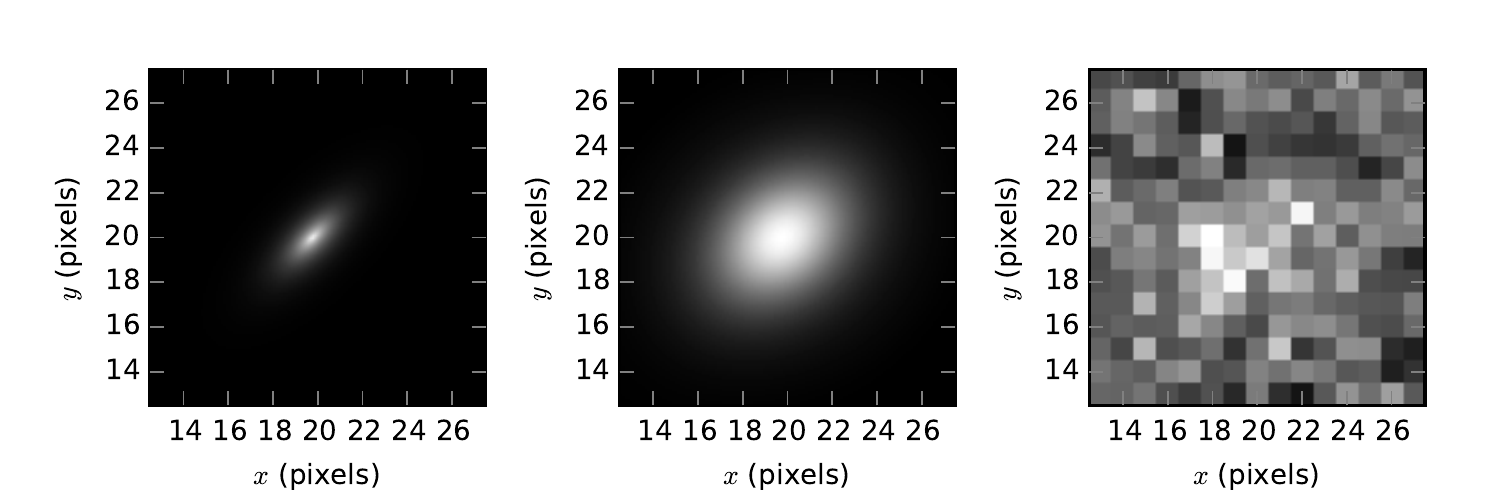}
\caption{An example of the simulated galaxy images used during the tests. Here, we have focused on the central regions of the images. In each case the axis labels correspond to the pixels of the low resolution image, and the colour scales are matched to the minimum and maximum pixel values in each image for demonstration purposes. The \emph{left panel} shows the source galaxy before PSF convolution and adding noise ($I_{\mathrm{gal}}$). The \emph{middle panel} shows the source convolved with the PSF ($I_{\mathrm{gal}}^*$). The final image ($I^{\mathrm{obs}}_{\mathrm{gal}}$) is pixelated and contains noise, and is shown in the \emph{right panel}.}
\label{fig:gal_im}
\end{minipage}
\end{figure*}

\subsection{Measuring position angles from simple simulations}
\label{subsec:demo}

To demonstrate an application of the RPC method when measuring position angles from realistic galaxy images, we simulate sources using the publicly available galaxy simulation software {\tt{GALSIM}}\footnote{\url{https://github.com/GalSim-developers/GalSim}} \citep{rowe14}. The images consist of $39\,\mathrm{pixel}\times39\,\mathrm{pixel}$ postage stamps, with a source placed close to the centre of each stamp.

The sources are assumed to have an exponential profile:
\begin{equation}
I_{\mathrm{gal}}\left(\theta\right)=I_0\exp{\left({-\frac{\theta}{\theta_d}}\right)},
\end{equation}
where $I_0$ is the peak intensity, the scale parameter $\theta_d=r_{1/2}/1.678$ and $r_{1/2}$ is the half-light radius. An exponential profile is known to provide a good description of the profiles of disc-dominated galaxies, which make up a large fraction of weak lensing sources.

We convolve the sources with a PSF described by a Moffat profile:
\begin{equation}
I_{\mathrm{PSF}}\left(\theta\right)=\frac{\beta-1}{\pi\alpha^2}\left[1+\left(\frac{\theta}{\alpha}\right)^2\right]^{-\beta},
\end{equation}
where we set $\alpha=3.01$ and $\beta=2.5$. These values of $\alpha$ and $\beta$ imply that the FWHM of the PSF is $3.4\,\mathrm{pixels}$. The width of the PSF in terms of pixel scale was chosen to approximately match the median PSF width for DES \citep{zuntz18}. The camera used for DES is DECam\footnote{\url{http://www.ctio.noao.edu/noao/node/1033}}, which has a pixel resolution of $\sim$$0.26\,\mathrm{arcsecs}$. Assuming this pixel resolution, the FWHM of our simulated PSF is $R_{\mathrm{p}}=0.9\,\mathrm{arcsecs}$, which is approximately the median seeing for DES.

The final simulated observed pixelated galaxy image, $I^{\mathrm{obs}}_{\mathrm{gal}}$, also contains noise, with
\begin{equation}
I^{\mathrm{obs}}_{\mathrm{gal}}\left(\bm{\theta}_{ij}\right)=I_{\mathrm{gal}}^{*}\left(\bm{\theta}_{ij}\right)+N\left(\bm{\theta_{ij}}\right),
\end{equation}
where the subscripts $i,j$ are the pixel indices, the asterisk denotes the pixelated, PSF-convolved source (before adding pixel noise) and $N(\bm{\theta}_{ij})$ is the pixel noise. In Fig. \ref{fig:gal_im}, we show subsequent stages when creating a simulated galaxy image.

An image of the PSF is also output from {\tt{GALSIM}} at the same pixel resolution as the galaxy images.

To measure the position angle of a source from the simulations, we first rotate the PSF image about the centroid by $90^{\circ}$ and then convolve the rotated PSF with the observed galaxy image, so that the image from which we measure the position angle using the RPC method can be written as
\begin{equation}\label{eq:final_im}
\bar{I}\left(\bm{\theta}\right)=\int\mathrm{d}^2\bm{\theta'}I^{\mathrm{obs}}_{\mathrm{gal}}\left(\bm{\theta}'\right)\hat{I}^{\perp}_{\mathrm{PSF}}\left(\bm{\theta}-\bm{\theta}'\right),
\end{equation}
where the rotated PSF image is
\begin{equation}
\hat{I}^{\perp}_{\mathrm{PSF}}\left(x,y\right)=I_{\mathrm{PSF}}\left(y,-x\right)+N_{\mathrm{PSF}}\left(y,-x\right),
\end{equation}
with $N_{\mathrm{PSF}}$ representing pixel noise on the PSF image.

The PSF at the position of a galaxy can be estimated by measuring the shapes of nearby stars and interpolating between their shapes. For this work we assume that all uncertainties in the PSF image can be modelled as Gaussian pixel noise. This approach ignores some of the systemic effects that can contaminate PSF estimation, such as the fatter-brighter effect (\citealt{antilogus14,guyonnet15,gruen15,mandelbaum18a}) whereby the PSF for an object increases with flux. A detailed look at the effects of these kinds of systematics on angle-only weak lensing will be carried out in future work.

In order to minimise the effects of noise at large scales, we apply an isotropic Gaussian weight function, $W(\bm{\theta})$, to the image $\bar{I}\left(\bm{\theta}\right)$, with the width of $W(\bm{\theta})$ selected to maximise signal-to-noise. This is achieved by maximising the function
\begin{equation}\label{eq:max_snr}
f\left(\sigma_{\mathrm{W}}\right)=\frac{\sum_{ij}{W(\bm{\theta}_{ij},\sigma_{\mathrm{W}})\bar{I}\left(\bm{\theta}_{ij}\right)}}{\sqrt{\sum_{ij}W^2(\bm{\theta}_{ij},\sigma_{\mathrm{W}})}},
\end{equation}
where the Gaussian weight function is written explicitly in terms of the width parameter $\sigma_{\mathrm{W}}$. The numerator in equation (\ref{eq:max_snr}) is the total flux after the weight function is applied. The denominator is the standard deviation of the noise after the weight function is applied, assuming the pixel noise has unity dispersion.

The centroid of each simulated source is estimated in unison with the fitting of the weight function. This is achieved by first convolving the image with the rotated PSF and fitting the weight function centred on the brightest pixel of the convolved image. We then update our estimate of the centroid using first-order moments and refit the weight function centred on the updated estimate. This last step is iterated until the difference between subsequent estimates of the centroid is less than 1\%.

We then measure the second-order moments, $\bar{Q}_{ij}$, of the weighted profile:
\begin{equation}
\bar{Q}_{MN}=\sum_{ij}\left(x_i-\bar{x}\right)^M\left(y_j-\bar{y}\right)^NW\left(\bm{\theta}_{ij}\right)\bar{I}\left(\bm{\theta}_{ij}\right).
\end{equation}
Finally, the position angle of the galaxy, $\alpha$, is estimated as
\begin{equation}\label{eq:alpha_uv}
\hat{\alpha}=\frac{1}{2}\tan^{-1}\left(\frac{\hat{v}}{\hat{u}}\right),
\end{equation}
where we have defined the Stokes parameters
\begin{align}\label{eq:stokes_params}
\hat{u}=&\bar{Q}_{20}-\bar{Q}_{02},\nonumber\\
\hat{v}=&2\bar{Q}_{11}.
\end{align}

Initial tests of this method are carried out without pixel noise and with the sources placed at the centre of each stamp. In anticipation of tests discussed later, we still apply an isotropic Gaussian weight function when recovering the position angles and also estimate the centroids from the images.

We simulate sources with ellipticity $|\epsilon|=0.384$. The sources are convolved with an elliptical PSF, as described above. In these tests, we set the half-light radius of the source galaxies to $r_{1/2}=1.3$. This corresponds to a ratio of $R_{\mathrm{gp}}/R_{\mathrm{p}}=1.3$, where $R_{\mathrm{gp}}$ is the FWHM of the PSF-convolved source and $R_{\mathrm{p}}$ is the FWHM of the PSF. We note that this value of  $R_{\mathrm{gp}}/R_{\mathrm{p}}$ is at the low end of the distribution for sources used to measure shapes with {\tt IM3SHAPE} for the DES Y1 shape catalogue \citep{zuntz18}; however, we show that we can still recover reliable position angles from these sources.

\begin{figure*}
\begin{minipage}{6in}
\centering
\includegraphics{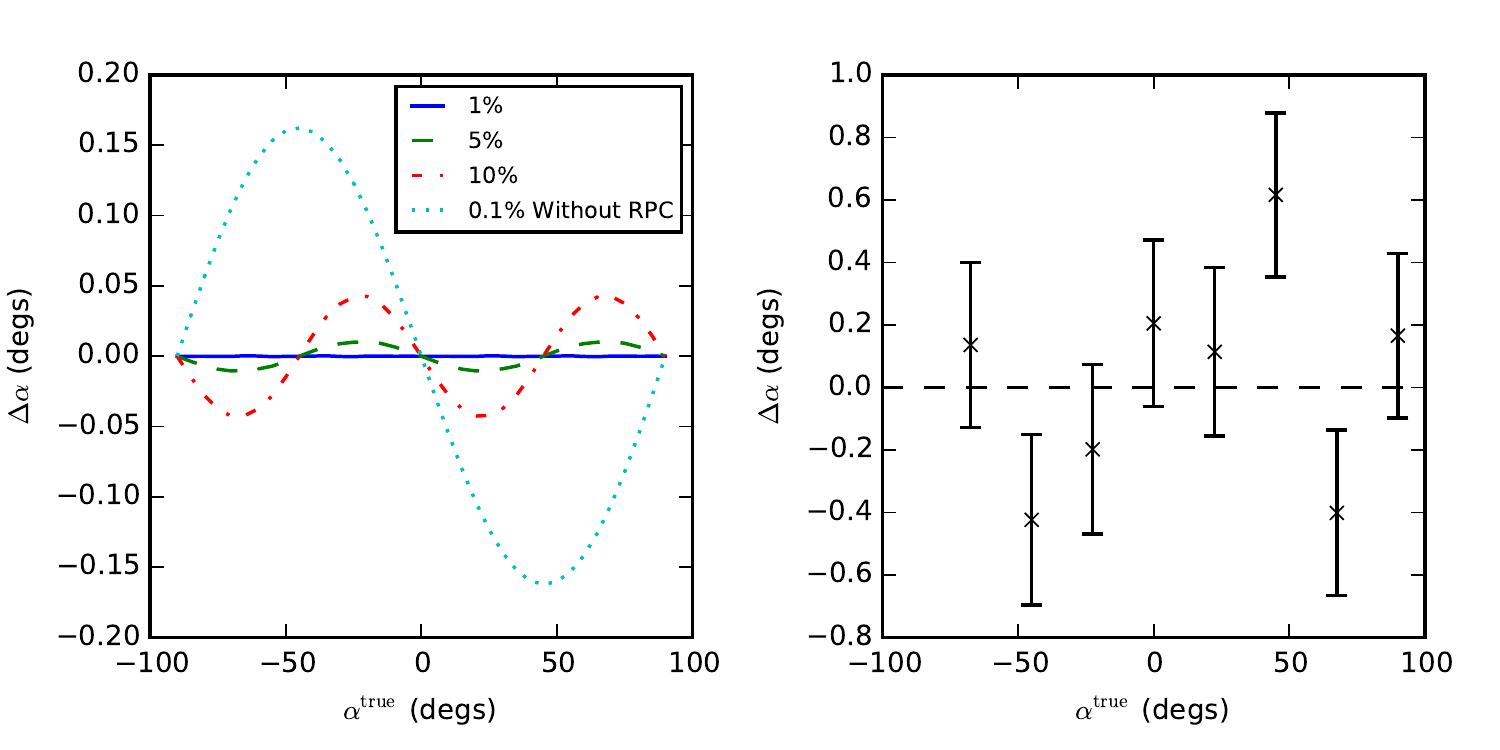}
\caption{\emph{Left panel}: Bias in the position angles measured from noise free simulated sources using the RPC method as a function of input position angle, and for three different PSF ellipticities: 1\%, 5\%, and 10\%. The dotted cyan curve shows the case for a PSF with 0.1\% ellipticity but where we do not use RPC, and so measure the position angles directly from the simulated sources. \emph{Right panel}: The mean estimated bias from $10^4$ noisy simulated sources with the PSF fixed at 5\% ellipticity. The black-dashed line shows zero bias.}
\label{fig:alpha_orient}
\end{minipage}
\end{figure*} 

We begin by convolving the sources with a PSF with 1\% ellipticity and position angle $0.0^{\circ}$. We create noise free images of sources with position angles in the range $-90^{\circ}<\alpha^{\mathrm{true}}<90^{\circ}$, and the position angles are recovered from these images using the RPC method, as discussed above. The results of this test are shown as the solid-blue curve in the left-hand panel of Fig. \ref{fig:alpha_orient}. This test is repeated for PSF ellipticities of 5\% and 10\%, and these are also shown in the left-hand panel of Fig. \ref{fig:alpha_orient}. As expected, the residual bias gradually increases with PSF ellipticity; however, the bias is very small - less than $0.05^{\circ}$- in all cases. It should be pointed out that no upsampling of the PSF or galaxy images is performed to achieve these results, but one might expect upsampling to reduce the residual bias further. Due to the small biases found in these tests, the effect of upsampling is not investigated here but will be considered in future work. As a comparison, we also include position angles measured using moments calculated directly from  $I_{\mathrm{gal}}^{\mathrm{obs}}$ (i.e. without correcting for the PSF using RPC) for the case of 0.1\% PSF ellipticity; this is shown as the dotted cyan curve. From this, we clearly see the benefit of the convolution step.

Next we add noise to the simulations but fix the PSF ellipticity to 5\%. The signal-to-noise-ratio (SNR) of the simulated sources is set to SNR=15, with the definition of SNR being that used by \cite{bridle10} and for the DES Y1 shape catalogue:
\begin{equation}
\mathrm{SNR}=\frac{\sqrt{\sum_{ij}\left[I_{\mathrm{gal}}^{*}\left(\bm{\theta_{ij}}\right)\right]^2}}{\sigma_{\mathrm{I}}}.
\end{equation}
where $\sigma_{\mathrm{I}}$ denotes the dispersion of pixel noise, which is assumed to be Gaussian distributed with $\sigma_{\mathrm{I}}=1$ and zero mean. It should be pointed out that this signal-to-noise is low compared with the majority of sources analysed by {\tt IM3SHAPE} for the DES Y1 shape catalogue.

For the tests that follow, we assume that we have an accurate image of the PSF at the positions of the sources and model uncertainties by adding Gaussian pixel noise to the PSF images from {\tt GALSIM} with $\mathrm{SNR}=50$. 
 
We again vary the position angles of the input sources and produce $10^4$ noise realisations for each position angle selected. The centroids of the sources were also allowed to vary uniformly within the central pixels of the postage stamps. As discussed above, the centroids are estimated from the sources assuming no prior knowledge of their values.
 
 The results of these tests are shown in the right-hand panel of Fig. \ref{fig:alpha_orient}. From this plot we see no clear evidence of a bias at a level of $<$$1^{\circ}$.

\subsection{Errors on position angle estimates}

\begin{figure*}
\begin{minipage}{6in}
\centering
\includegraphics{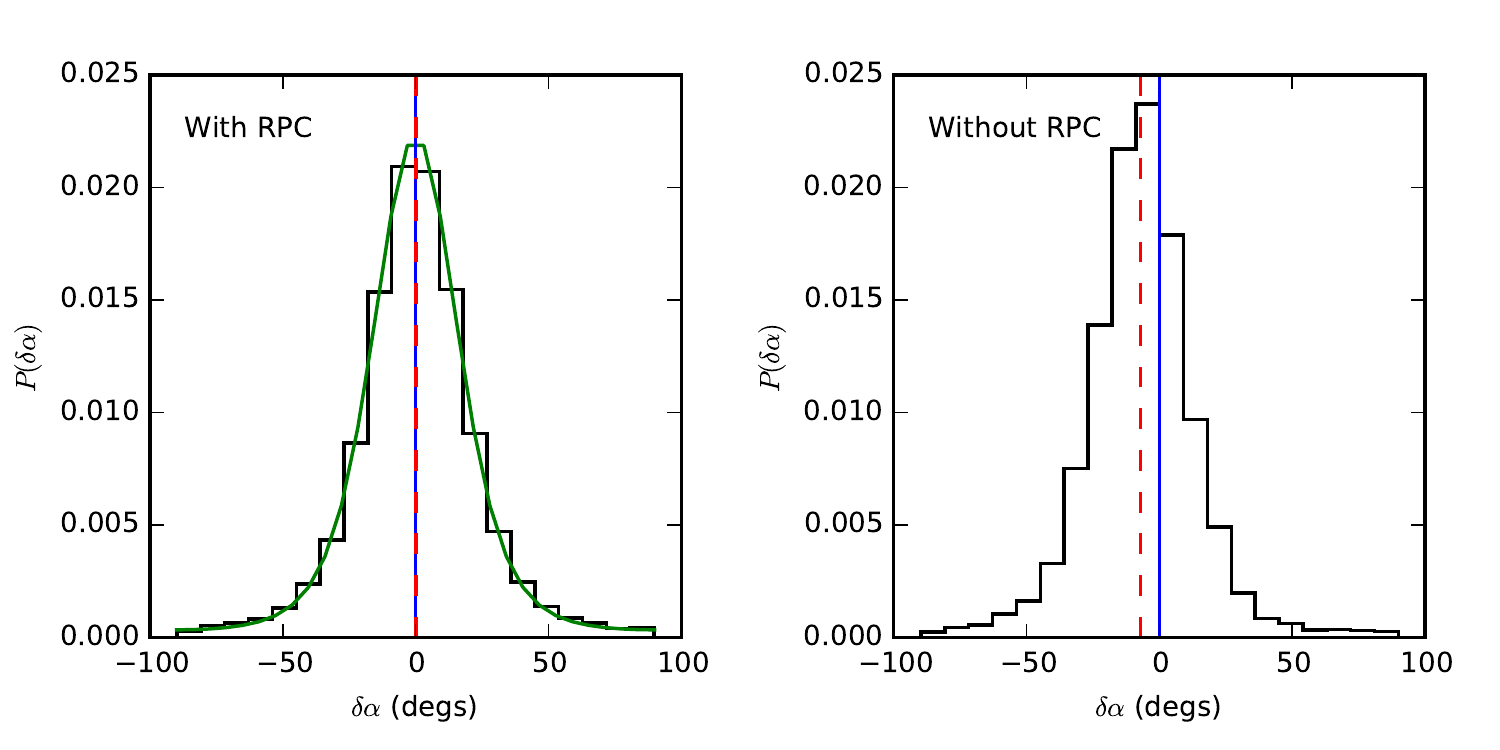}
\caption{Distribution of $\delta\alpha=\hat{\alpha}-\alpha^{\mathrm{true}}$ for $10^4$ noise realisations. The \emph{left panel} shows the results using the RPC method to correct the observed images for PSF convolution. The green curve shows the distribution of $\delta\alpha$ predicted using equation (\ref{eq:marg_chi}). We see that equation (\ref{eq:marg_chi}) accurately describes the distribution of position angle estimates. The \emph{right panel} shows the case where we do not use RPC and so do not correct for the PSF. In both cases, the blue line indicates $\delta\alpha=0$ and the red-dashed line is the mean recovered estimate. It is clear that the RPC method has significantly reduced the effects of PSF convolution.}
\label{fig:dalpha}
\end{minipage}
\end{figure*} 

One of the attractive properties of using moments and RPC to estimate the position angles is that it is possible to analytically calculate what the distribution of estimates should be if we assume Gaussian distributed pixel noise on both the galaxy image and the PSF image.

If we assume that we know the centroid, then for a given weight function, the distribution of estimated position angles is shown in Appendix \ref{ap:err_posang} to be a projected normal distribution \citep{mardia00}, with the functional form given in equation (\ref{eq:marg_chi}). A more thorough derivation that includes errors on the centroids is not considered, but the distribution of errors on the centroids will be specific to the method used to estimate them.

In order to investigate the distribution of position angle estimates using simulations, we repeat the simulations used for the right-hand side of Fig. \ref{fig:alpha_orient} for the case where $\alpha^{\mathrm{true}}=45.0^{\circ}$. For this test, we use a fixed-size Gaussian weight function, with full-width half-maximum $R_w=(R_g^2+2R^2_p)^{1/2}$, where $R_g$ is the full-width half-maximum of the galaxy. The factor of two is included because each source is convolved twice with the PSF: once when creating the observed image and once when estimating the position angle. We fix the size of the weight function so that we can compare the distribution of position angle estimates with the distribution given in equation (\ref{eq:marg_chi}). This distribution does not account for errors on the centroids; nevertheless, in these tests, we estimate centroids from the images.

The results of this test are shown in the left-hand panel of Fig. \ref{fig:dalpha}. The mean recovered estimate is found to be $<\hat{\alpha}>=(45.2\pm0.2)^{\circ}$, which is consistent with zero bias. The right-hand
panel shows the position angles measured from the same simulations without using RPC, as a comparison. The mean recovered estimate in this case is $<\hat{\alpha}>=(37.2\pm0.2)^{\circ}$. We find a small 12\% increase in errors when using RPC; however, we emphasise that the ratio of $R_{\mathrm{gp}}/R_{\mathrm{p}}=1.3$ is at the low end for sources used in the DES Y1 {\tt IM3SHAPE} analysis, and one would expect the increase in errors to be smaller for sources with larger $R_{\mathrm{gp}}/R_{\mathrm{p}}$.

The distribution given in equation (\ref{eq:marg_chi}) is shown for the simulations discussed above as the green curve in the left-hand panel of Fig. (\ref{fig:dalpha}). Equation (\ref{eq:marg_chi}) assumes that errors on centroids (which were estimated from the images in the simulations) are negligible; however, we see that, even with this assumption, equation (\ref{eq:marg_chi}) accurately describes the distribution of estimates.

In the discussion following equation (\ref{eq:marg_chi}), we suggest that some residual bias in the position angle estimates is expected when considering elliptical PSFs. This bias is small enough to be undetected in the simulations above. A small PSF dependent bias is detected in the shear estimates when we analyse the Stage III simulation (see Section \ref{subsec:rec_shear}); however, in Section \ref{sec:results} we show that the level of bias does not have a significant impact on the inferred cosmology. This may be a problem for Stage IV weak lensing, but an investigation of this will be carried out in future work.

\section{Simulated weak lensing survey}
\label{sec:cos_sim}
We test the angle-only shear estimator on a simulated Stage III weak lensing set of galaxy images, with galaxy numbers and survey area chosen to be similar to those expected for the full DES. The shears are estimated using equation (\ref{eq:linest}), and the position angles are measured using the RPC method discussed in the previous section. The goal of the paper is to show that we can recover input cosmology using angle-only shear estimates. The details of the simulation are discussed in this section, with the details of the shear estimates and cosmological inference presented in the following sections.

We begin by simulating full-sky Gaussian random shear fields for four redshift bins using {\tt HEALPIX}\footnote{\url{https://healpix.sourceforge.io}} \citep{gorski05} with a resolution of $N_{\mathrm{side}}=1024$. From these full-sky maps, we extract pixels in circular regions around RA = $0.0^{\circ}$, Dec = $0.0^{\circ}$ with areas of $5,000\,\mathrm{deg}^2$. Each extracted region contains $\sim$1.52 million pixels, and these maps provide the input shears for the simulated survey used for the remainder of this paper. A circular region is selected in order to minimise the generation of B-modes in the cut-sky power spectra, since the effects of a cut sky are not the focus of this paper. We will estimate the power spectra up to multipole $l_{\mathrm{max}}$ = 2048, and so we include theory power spectra up to $l_{\mathrm{sim}}$ = 3071 in the {\tt HEALPIX} maps to allow for multipole mixing from higher modes.

It should be emphasised that accurate forecasting for DES is not the aim of this paper; we wish to show that angle-only shear shear estimates can be used to recover cosmology from a Stage III-like experiment. Therefore, the details of the simulation are only intended to loosely match those of DES. 

The cosmology assumed for the shear fields is a $\Lambda\mathrm{CDM}$ model with the TT,TE,EE+lowE+lensing+BAO Planck 2015 best-fit parameters given in \cite{Ade16}. The input values of the parameters fitted for in this paper are $\Omega_{\mathrm{m}}=0.3089$ and $\sigma_8=0.8159$, and $w=-1$.

\begin{figure}
\includegraphics{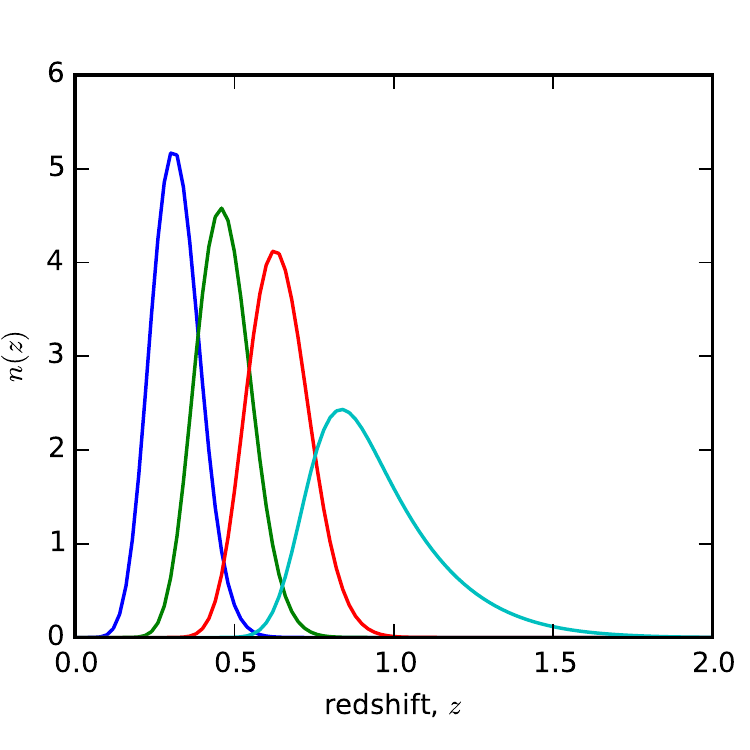}
\caption{The four overlapping redshift bins used in the simulation. The overlap between the bins is due to photo-$z$ errors causing some sources to be placed in the wrong bins.}
\label{fig:nz}
\end{figure} 

The redshift distribution of the simulated sources is assumed to be \citep{smail94}
\begin{equation}
n\left(z\right)\propto z^2\exp\left[-\left(\frac{z}{\bar{z}}\right)^{\beta}\right],
\end{equation}
with $\beta=1.5$, $\bar{z}=0.39$ and minimum redshift $z_{\mathrm{min}}=0.2$. This corresponds to a median redshift of $z_{\mathrm{med}}=0.59$. The $z$-bin widths are selected so that the number of sources in each bin is the same. A photo-$z$ error is included, with dispersion $\sigma_{\mathrm{phz}}=0.05(1+z)$. The redshift distribution of the sources within each of the four bins is shown in Fig. \ref{fig:nz}.

For each pixel in the four shear maps, 50 source galaxies are simulated, providing a total number of $\sim$305 million sources at a source density of $\sim$$17\,\mathrm{arcmin}^{-2}$. For each source, a $39\,\mathrm{pixel}\times39\,\mathrm{pixel}$ postage stamp image is created assuming an exponential intensity profile (as discussed in Section \ref{sec:meas}), and the centroids are randomly selected to lie anywhere within the central pixel.

\begin{figure*}
\begin{minipage}{6in}
\centering
\includegraphics{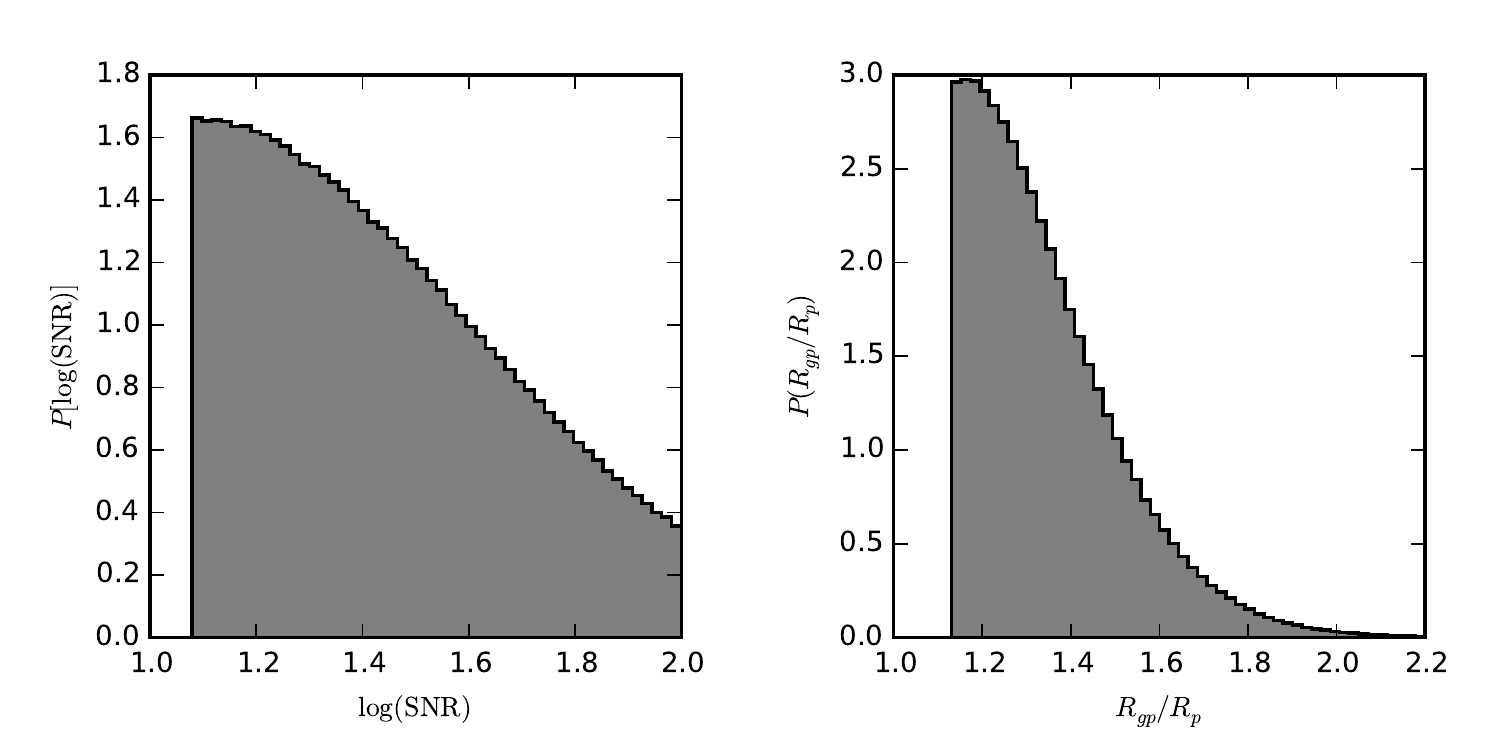}
\caption{Distributions of SNR (\emph{left panel}) and $R_{gp}/R_p$ (\emph{right panel}) for 1\% of the sources randomly selected from the distributions used in the Stage III-like simulation.}
\label{fig:SNR}
\end{minipage}
\end{figure*}

Samples of SNR are drawn from the lognormal distribution
\begin{equation}\label{eq:SNR_lognorm}
f\left(\mathrm{SNR}\right)\propto\frac{1}{\mathrm{SNR}}\exp\left[-\frac{\left(\ln\mathrm{SNR}-\mu_{\mathrm{SNR}}\right)^2}{2\sigma_{\mathrm{SNR}}^2}\right],
\end{equation}
where values of $\mu_{\mathrm{SNR}}=2.56$ and $\sigma_{\mathrm{SNR}}=1.1$ are selected so that the distribution is qualitatively similar to that of the sources used in the DES Y1 {\tt IM3SHAPE} analysis \citep{zuntz18}. The distribution is shown in the left-hand panel of Fig. \ref{fig:SNR}.

The half-light radii of the sources are selected so that the distribution of $R_{\mathrm{gp}}/R_{\mathrm{p}}$ is qualitatively similar to that in the DES Y1 {\tt IM3SHAPE} analysis. To achieve this, we draw samples of $x_{\mathrm{R}}=R_{\mathrm{gp}}/R_{\mathrm{p}}$ from the lognormal distribution shown in the right-hand panel of Fig. \ref{fig:SNR}:
\begin{equation}\label{eq:Re_lognorm}
f\left(x_{\mathrm{R}}\right)\propto\frac{1}{x_{\mathrm{R}}}\exp\left[-\frac{\left(\ln x_{\mathrm{R}}-\mu_{\mathrm{R}}\right)^2}{2\sigma_{\mathrm{R}}^2}\right],
\end{equation}
with parameters $\mu_{\mathrm{R}}=0.18$ and $\sigma_{\mathrm{R}}=0.18$.

In order to convert sampled values of $R_{\mathrm{gp}}/R_{\mathrm{p}}$ to half-light radii for use as inputs to {\tt GALSIM}, we use {\tt GALSIM} to produce images of circular, noise free, PSF-convolved sources for a selection of input half-light radii, in increments of $\Delta r_{1/2}=0.25\,\mathrm{pixels}$, and measure the $R_{\mathrm{gp}}/R_{\mathrm{p}}$ of these images. This gives us a relationship between $r_{1/2}$ and $R_{\mathrm{gp}}/R_{\mathrm{p}}$. For each random sample of $R_{\mathrm{gp}}/R_{\mathrm{p}}$ drawn from the distribution given in equation (\ref{eq:Re_lognorm}), we interpolate into this relationship to give us a sample $r_{1/2}$ which is then used in the simulation. The relationship between $r_{1/2}$ and $R_{\mathrm{gp}}/R_{\mathrm{p}}$ is found to be approximately linear in the assumed region $1.13\le R_{\mathrm{gp}}/R_{\mathrm{p}}\le2.2$, with $r_{1/2}=m(R_{\mathrm{gp}}/R_{\mathrm{p}})+c$, where $m=3.12$ and $c=-2.91$.

The 1D components of the intrinsic ellipticities of the sources, $\epsilon_i^{\mathrm{int}}$, are assumed to be Gaussian distributed, with zero mean, 1D dispersion parameter $\sigma=0.29$, and range $0\le|\epsilon^{\mathrm{int}}|\le0.804$. The maximum value of $|\epsilon^{\mathrm{int}}|$ is selected to match that found in the CFHTLenS {\tt LENSFIT} analysis \citep{miller13}. Placing this cut on $|\epsilon^{\mathrm{int}}|$ slightly modifies the 1D dispersion of the ellipticities as measured from the simulated data, so that the intrinsic 1D shape dispersion in the simulation is $\sigma_{\epsilon}=0.278$.

\begin{figure*}
\begin{minipage}{6in}
\includegraphics{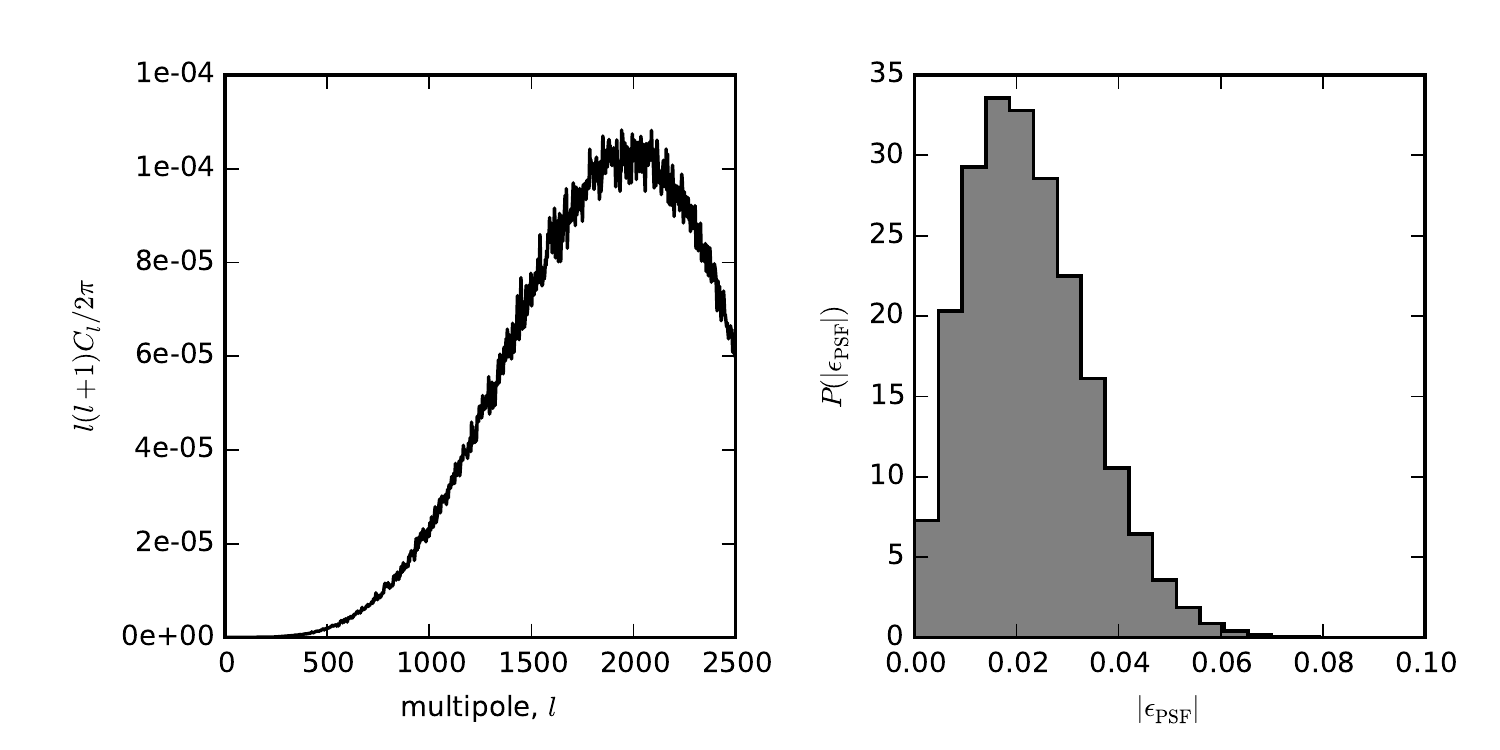}
\caption{The pseudo-$C_l$s of the PSF ellipticities used for the Stage III-like simulation are shown in the \emph{left panel}. The distribution of PSF ellipticities is given in the \emph{right panel}.}
\label{fig:psf_cls}
\end{minipage} 
\end{figure*}

The sources are convolved with a PSF modelled as a Moffat profile, with $\alpha=3.01$, $\beta=2.5$, and the FWHM fixed to 3.4 pixels, as discussed in Section \ref{sec:meas}. The ellipticity of the PSF is allowed to vary over the observed region of sky. To model this variation, we first create {\tt HEALPIX} maps for the two components of PSF ellipticity using the toy input power spectrum
\begin{equation}
C_l\propto1-\cos\left(\frac{2\pi l}{l_{\mathrm{sim}}}\right).
\end{equation}
The selection of this power spectrum is somewhat arbitrary, but it provides scale-dependent correlations across the whole range of multipole modes.

We extract PSF ellipticities from the maps for the same pixels used to create the shear maps when simulating galaxy shapes. The ellipticities are rescaled to ensure that the maximum PSF ellipticity is $|\epsilon_{\mathrm{PSF}}|=0.1$, and we use these PSF ellipticity maps for each of the four $z$-bins. In the left-hand panel of Fig. \ref{fig:psf_cls}, we show the pseudo-$C_l$ power spectrum calculated from the extracted, rescaled PSF ellipticities using the {\tt HEALPIX} subroutine {\tt anafast}. In the right-hand panel of Fig. \ref{fig:psf_cls}, we show the distribution of $|\epsilon_{\mathrm{PSF}}|$.

We provide noisy, independent PSF images for each source so that correlations in PSF uncertainty are not propagated into the shear power spectra recovered in Section \ref{sec:ps}. The aim is to show that cosmology can be accurately inferred from PSF-convolved galaxy images using only position angle estimates if one has an accurate estimate of the PSF. To this end, uncertainties in the PSF at the position of each galaxy are modelled purely as Gaussian random pixel noise, with SNR = 50. Systematics that may arise from uncertainties in the PSF model are left for future work.

The noisy, PSF-convolved galaxy images and the noisy PSF images provide the simulated observed data set used for the remainder of this paper.

\section{Estimating the shear signal}
\label{sec:est_shear}
In this section, we discuss the methods used to recover shear from the simulated data set. We assume the angle-only shear estimator in equation (\ref{eq:linest}), and we investigate the estimates for multiplicative and additive biases.

\subsection{Recovering the shear}
\label{subsec:rec_shear}

We begin by estimating the position angles of the simulated galaxy images using quadrupole moments and the RPC method, as discussed in section \ref{subsec:demo}. We also estimate the centroids and select the width of the weight function so that the signal-to-noise is maximised. From the position angles, we calculate the mean $\cos(2\hat{\alpha})$ and mean $\sin(2\hat{\alpha})$ in each pixel of the observed regions for the four $z$-bins. This provides us with two$\times$four additional maps: a map of mean cosines, $C_i(\bm{\theta})$, and a map of mean sines,  $S_i(\bm{\theta})$, for each $z$-bin denoted by subscript $i$. 

To relate means of cosines and sines to shear estimates, we require an estimate of the multiplicative factor $\mu$ in equation (\ref{eq:linest}). To estimate $\mu$, we create a calibration data set by repeating the simulation used to create the observed data set, but with the magnitude of the shears fixed to $|g|=0.01$. A random shear position angle, $\alpha_0$, is assigned to the shear in each pixel. All other properties of the simulations are kept the same, so that there are $N_{\mathrm{sim}}\sim$305 million galaxies in the calibration data set.

Letting $\xi=1/\mu$, we then calculate
\begin{equation}\label{eq:m_est}
\hat{\xi}=\frac{1}{N_{\mathrm{sim}}}\sum_{k=1}^{N_{\mathrm{sim}}}\frac{\cos\left(2\hat{\alpha}^{(k)}-2\alpha_0^{(k)}\right)}{\left|g\right|},
\end{equation}
where the summation is over all the calibration sources and $\alpha_0^{(k)}$ is the position angle of the shear on the $k^{\mathrm{th}}$ source. This approach to measuring $\mu$ using a simulation with known input shear values is similar to approaches commonly used by shape measurement techniques to calibrate for multiplicative biases. The error on $\hat{\xi}$ is expected to be Gaussian distributed due to the central limit theorem. From this simulation, we find $\hat{\xi}=1.743\pm0.004$. Since the error is much smaller than $\xi$, the error on $\hat{\mu}=1/\hat{\xi}$ will also be approximately Gaussian, with $<\hat{\mu}>\approx<1/\hat{\xi}>$. Hence, the factor $\mu$ is estimated as $\hat{\mu}=0.574\pm0.001$.

For the more general case, where the $F_1$ function is not accurately described by the first-order approximation given in equation (\ref{eq:approx_F}), we may be required to repeat these simulations for a range of input $|g|$ values so that we can accurately sample the $F_1$ function over that range. We could then interpolate into these samples to provide estimates of $|g|$. This will likely be required when we look at more sophisticated Stage IV simulations in future work, and so an investigation of this will be carried out then.

An expression for $\mu$ when measurement errors can be considered negligible is given in equation (\ref{eq:taylor_k}). A general expression for $\mu$ that includes measurement error information is difficult to derive. For the simulation discussed in Section \ref{sec:cos_sim}, we include distributions of galaxy sizes and signal-to-noise ratios (SNR) (these contribute to vector $\bm{D}$ in equation (\ref{eq:F1_errdist})), and we assume that these distributions are the same at all angular positions on the sky and for each redshift bin. This assumption allows us to produce a calibration simulation using shears with a constant magnitude from which we can estimate a single value of $\mu$. In a real survey, of course, these distributions would be redshift dependent, and so one would have to calculate values of $\mu$ for each redshift bin. What is potentially more problematic is the impact of angular position dependent errors (and $f(|\epsilon^{\mathrm{int}}|)$), which would modify the $F_1$ function across the sky. However, if we assume equation (\ref{eq:linest}), estimates of the shear are linearly dependent on $\mu$, and so if errors on $\mu$ are uncorrelated with the shear and centred on zero, these uncertainties may be modelled for by including them in the likelihood when fitting for cosmology. In this case, it may still be possible to use a single $\mu$ for each redshift bin. In Section \ref{sec:likelihood}, we model for errors on our estimated $\mu$ by including the uncertainty in the covariance matrix. If one expects the value of $\mu$ to vary across the sky, it may also be possible to include information about this variation in the likelihood.

In the more general case, where we require the full $F_1$ function, variations in the shape of $F_1$ across the sky may be more difficult to account for, but the degree to which this is a problem is beyond the scope of this paper and will be investigated in future work.

We can also calculate an equivalent statistic to that in equation (\ref{eq:m_est}) for sine:
\begin{equation}\label{eq:mdash_est}
\hat{\xi}'=\frac{1}{N_{\mathrm{sim}}}\sum_{k=1}^{N_{\mathrm{sim}}}\frac{\sin\left(2\hat{\alpha}^{(k)}-2\alpha_0^{(k)}\right)}{\left|g\right|}.
\end{equation}
If this statistic is not consistent with zero, there are non-zero odd moments of $\hat{\alpha}-\alpha_0$, and so the distribution of $\delta$ is skewed. In this case, there will be a systematic rotation of the shear estimates. From the calibration simulation used to estimate $\hat{\xi}$, we find $\hat{\xi}'=(7.93\pm4.05)\times10^{-3}$. This may indicate a small skewness in the distribution of $\delta$, although it is not clear what the source of such a skewness could be, since the $\alpha_0^{(i)}$ are drawn from a uniform distribution. From these results, we can draw no conclusion on whether a skewness exists in the calibration simulation, but $\xi'\ll\xi$, and a similar test applied to the position angles from the simulated observed data set using the known input shear position angles returned a value of $\xi'$ consistent with zero; therefore, we do not consider this to be a problem.

Finally, maps of the shear estimates are created for each $z$-bin using the means of the trigonometric functions, where
\begin{align}\label{eq:linest_maps}
\hat{g}_1^{(i)}\left(\bm{\theta}\right)=&\hat{\mu}C_i\left(\bm{\theta}\right),\nonumber\\
\hat{g}_2^{(i)}\left(\bm{\theta}\right)=&\hat{\mu}S_i\left(\bm{\theta}\right).
\end{align}

\subsection{Performance of the estimator}
\label{subsec:shear_performance}
\begin{figure*}
\begin{minipage}{6in}
\centering
\includegraphics{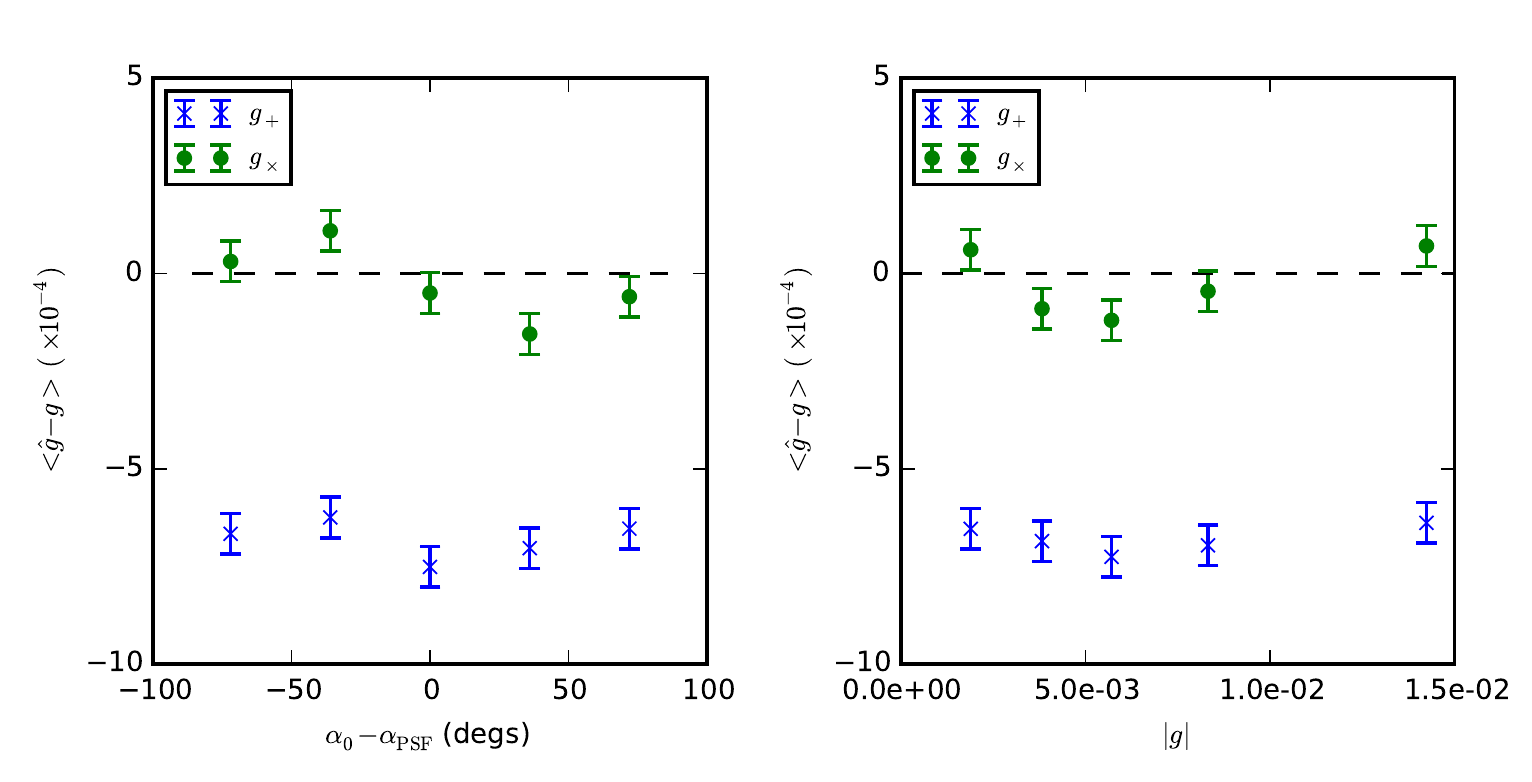}
\caption{Bias in the shears estimated from the simulation as a function of the difference between the input shear and PSF position angles (\emph{left panel}) and the modulus of the input shear (\emph{right panel}). There is no clear evidence of a bias that depends on either of the input quantities and any absolute bias is below $10^{-4}$.}
\label{fig:shear_bias}
\end{minipage}
\end{figure*}

We investigate the performance of the estimator by comparing the recovered shears with the input shears. For each estimated shear and corresponding input shear we calculate $\hat{g}-g$ in a coordinate system aligned with the major axis of the local PSF. This coordinate system is selected because we expect residual bias from imperfect PSF correction to be the dominant source of bias in our estimator. This choice of coordinate system was also adopted for the GREAT3 challenge \citep{mandelbaum14}.

We first bin the two components of $\hat{g}-g$  with respect to the difference between the local shear position angle and the local direction of the PSF, $\Delta\alpha_0=\alpha_0-\alpha_{\mathrm{PSF}}$, to search for systematics that depend on the alignment of the shear and PSF. The bins are constructed so that each bin contains the same number of galaxies: $\sim$61 million sources per bin. This is shown in the left-hand panel of Fig. \ref{fig:shear_bias}, where $+$ and $\times$ denote the components of the shear in the direction aligned with and at $45^{\circ}$ to the PSF respectively. Here, we see clear evidence of a bias in $g_+$ at a level of $\sim$$7\times10^{-5}$. There is no clear bias that depends on the direction of the shear with respect to the PSF orientation, but there is some evidence of sinusoidal behaviour in both components.

Next, we repeat the test above, but this time we bin with respect to the modulus of the shear (using the same number of sources in each bin) to look for shear magnitude dependent biases. This is shown in the right-hand panel of Fig. \ref{fig:shear_bias}. There is no evidence of a $|g|$ dependent bias. Here, the bias in $g_+$ is the same bias found in the left panel.

The overall bias for the full set of $\sim$$3\times10^8/50$ shear estimates (since we assume 50 sources per pixel) is\\
$<\hat{g}_+-g_+>=(-68.0\pm2.3)\times10^{-5}$ and\\$<\hat{g}_{\times}-g_{\times}>=(2.50\pm2.32)\times10^{-5}$.\\The bias in $g_+$ is expected to arise from an imperfect PSF correction.

\begin{table*}
\begin{minipage}{6in}
\centering
\begin{tabular}{|c|c|c|c|c|c|}
\hline
Method & $m_+$ ($\times10^{-3}$) & $m_{\times}$ ($\times10^{-3}$) & $c_+$ ($\times10^{-5}$) & $c_{\times}$ ($\times10^{-5}$) & $Q_{\mathrm{c}}$\\ [0.5ex]
\hline
With RPC & $-5.91\pm4.00$ & $8.56\pm4.00$ & $-68.0\pm2.3$ & $2.50\pm2.32$ & $372_{-143}^{+175}$ \\ [1ex]
Without RPC & $-1.25\pm4.01$ & $9.89\pm3.56$ & $524.1\pm2.3$ & $-3.06\pm2.07$ & $9.40_{-0.01}^{+0.01}$ \\ [1ex]
{\tt IM3SHAPE}: uncalibrated & $-34.4\pm3.4$ & $-33.1\pm3.4$ & $-14.9\pm2.0$ & $-2.62\pm1.95$ & $98.4_{-12.2}^{+12.9}$ \\ [1ex]
{\tt IM3SHAPE}: calibrated & $-0.63\pm3.5$ & $0.63\pm3.5$ & $-14.9\pm2.0$ & $-2.62\pm1.95$ & $310_{-28}^{+19}$ \\ [1ex]
{\tt IM3SHAPE}: angle-only & $6.04\pm3.54$ & $7.24\pm3.54$ & $-7.51\pm2.06$ & $-3.52\pm2.06$ & $487_{-237}^{+344}$ \\ [1ex]
\hline
\end{tabular}
\caption{Estimated multiplicative and additive biases for the two shear components in a coordinate system that aligns with the local PSF. In the last column, we also give the value of $Q_{\mathrm{c}}$.}
\label{table:m_and_c}
\end{minipage}
\end{table*}

Next, we fit for the linear bias model
\begin{equation}\label{eq:lin_bias}
\hat{g}_{\alpha}-g_{\alpha}=m_{\alpha} g_{\alpha}+c_{\alpha},
\end{equation}
where we have used $\alpha$ to denote a component of shear (+/$\times$). We perform a least squares fit of the model parameters $m_{\alpha}$ and $c_{\alpha}$ to the shear estimates and true input shears, whereby we include all of the shear estimates recovered from the simulation. The results from this fit are shown in the row labelled ``With RPC''  in Table \ref{table:m_and_c}. From these results, we see some evidence of a non-zero multiplicative bias. The additive bias indicated above is also evident. In Section \ref{sec:results}, when we look at the cosmology inferred from the simulation using these shear estimates, we do not find evidence of a bias in the constraints. However, this level of multiplicative bias will likely be problematic for future Stage IV weak lensing surveys (see \cite{massey13} for details), and so the angle-only method would need to be refined. It may be that assuming a fixed shear magnitude in the calibration simulation is insufficient for estimating $\mu$, since the dispersions of errors on the position angles depend on the magnitudes of galaxy ellipticities. It may also be true that we need to consider a higher-order approximation or full form for the $F_1$ function; however, an investigation of this is left for future work, when we will consider Stage IV missions.

For comparison with shear estimation methods applied to the GREAT3 simulations (see \cite{mandelbaum15}), we calculate $Q_{\mathrm{c}}$, which is the metric used to determine the performance of the shear estimates submitted to the GREAT3 challenge \citep{mandelbaum14}), This is also presented in Table \ref{table:m_and_c}. Here, we quote the mean $Q_{\mathrm{c}}$ from a simple set of simulations that assume uncorrelated Gaussian distributed $m$ and $c$ values using the results quoted in Table \ref{table:m_and_c} as the means and dispersions. The error bars are the 95\% confidence limits.

If we assume that the residual bias from an imperfect PSF correction is accurately described by the linear bias model, we can correct for the additive bias found in Table \ref{table:m_and_c} if we know the position angle of the PSF, $\alpha_{\mathrm{PSF}}$, at the positions of the sources. Let us write the shear estimate at position $\bm{\theta}$ in terms of the true shear, $g(\bm{\theta})$,
\begin{equation}
\hat{g}(\bm{\theta})=g(\bm{\theta})+c',
\end{equation}
where the (complex) additive bias $c'$ is a bias in the reference frame aligned with $\alpha_0=0$ and we have assumed zero multiplicative bias. For each source, we can rotate to a reference frame aligned with the PSF at the position of that source:
\begin{equation}
R(\bm{\theta})\hat{g}(\bm{\theta})=R(\bm{\theta})g(\bm{\theta})+c,
\end{equation}
where $c=R(\bm{\theta})c'$ and $R(\bm{\theta})=e^{-2i\alpha_{\mathrm{PSF}}(\bm{\theta})}$. The parameter $c$ is the additive bias in the frame of reference aligned with the local PSF. One can estimate this from the data by averaging over the measured $R(\bm{\theta})\hat{g}(\bm{\theta})$, since the expectation value of the true shear is zero. Upon doing this, we estimate $\Re(c)=(-68.0\pm2.3)\times10^{-5}$ and $\Im(c)=(-2.78\times2.34)\times10^{-5}$. Hence, the corrected estimates are
\begin{equation}\label{eq:cor_PSF}
\hat{g}^{\mathrm{cor}}(\bm{\theta})=\hat{g}-R^{-1}(\bm{\theta})c.
\end{equation}

Using the corrected shear estimates, we recalculate the additive and multiplicative biases and find $Q_{\mathrm{c}}=453^{+339}_{-230}$. Therefore, the $Q_{\mathrm{c}}$ slightly increases, but there is also a large increase in the error bars.

There may also be some additive bias due to pixelation. Upon averaging over the estimated shears without performing the rotation step above, we calculate $\Re(c')=(-6.06\pm2.33)\times10^{-5}$ and $\Im(c')=(4.84\times2.34)\times10^{-5}$. Therefore, there is some evidence of a bias in the pixel reference frame. However, this bias is small and not deemed to be problematic.

A spatially varying additive bias on the shear estimates, such as one that depends on the PSF, propagates as an additive bias on the shear power spectrum. This bias is given by the power spectrum of the additive bias \citep{kitching19}. For the remainder of this paper, we focus on the uncorrected shear estimates which do contain a PSF dependent additive bias, but as a test, we have compared cosmological constraints inferred using both the corrected (using equation (\ref{eq:cor_PSF})) and uncorrected shear estimates and found no significant differences between the two. When considering Stage IV weak lensing experiments in future work, these levels of bias may be important.

In order to look at the benefit of using RPC, we also recover angle-only shear estimates using position angles measured without RPC, and so without correcting for the PSF. In this case, the value of $\mu$ is different to that found above because the noise properties are different when we include the convolution step. From the calibration simulation discussed above, we estimate $\hat{\mu}=0.513\pm0.009$ when RPC is not included. The results from this test are shown in the row labelled ``Without RPC'' in Table \ref{table:m_and_c}. Here, we see that the additive bias $c_+$ is an order of magnitude larger than it is when we use RPC. This leads to a large drop in the value of $Q_{\mathrm{c}}$. If we use equation (\ref{eq:cor_PSF}) to correct these shear estimates for the PSF-dependent additive bias, we increase the value to $Q_{\mathrm{c}}=464^{+329}_{-223}$. However, we find that we cannot recover accurate power spectra using these corrected shears. This indicates that there is a spatially varying residual additive bias when we try to correct the angle-only shear estimates that do not use RPC for the PSF using this simple linear bias model. This residual bias likely depends on the ellipticity of the PSF.

As a comparison with the angle-only method, we also recover shear estimates from the galaxy images using {\tt IM3SHAPE}. For {\tt IM3SHAPE}, we use the lowest resolution settings, with upsampling=1, and we assume the correct galaxy and PSF profiles in the models fitted to the images. With these settings, we find that {\tt IM3SHAPE} processes $10^3$ galaxies, without bias calibration, in $\sim$360 seconds using a single thread. The cosines and sines of the $10^3$ galaxies are estimated in $\sim$36 seconds using RPC with quadrupole moments. The RPC method is coded purely in Python and again uses a single thread. Hence, we find we can process galaxies an order of magnitude faster using RPC. These results compare the processing speed of uncalibrated {\tt IM3SHAPE} shape measurements with measurements of the cosines and sines of the galaxy position angles (i.e. with no estimate of the factor $\mu$). The increase in speed when going from shapes to position angles is because the position angles are measured using quadrupole moments, as opposed to fitting for galaxy models when measuring ellipticities.

Shears are estimated from the {\tt IM3SHAPE} ellipticities using equation (\ref{eq:st_est}) with uniform weighting, $w_i=1$. A uniform weighting scheme is adopted both for simplicity and to provide a fair comparison with the angle-only shear estimates, since we have not yet looked at the impact of introducing a weighting scheme when calculating the means of the trigonometric functions in equation (\ref{eq:linest}). This will be investigated in future work, but an unbiased weighting scheme should be one that is independent of galaxy orientation, as long as the same weighting is used in the calibration simulations when calculating $\mu$.

The biases and corresponding value of $Q_{\mathrm{c}}$ for uncalibrated shear estimates recovered using {\tt IM3SHAPE} are given in Table \ref{table:m_and_c}. Here, we see clear multiplicative and additive biases. We emphasise that these results are found when using {\tt IM3SHAPE} with the lowest possible resolution settings; therefore, there is expected to be some contribution from model bias, and one should not read too much into these results. The main reason we use {\tt IM3SHAPE} is to compare constraining power with the angle-only method when we look at recovering cosmology in Section \ref{sec:results}.

When fitting for cosmology using {\tt IM3SHAPE} in Section \ref{sec:results}, the shears are calibrated using the mean of $m_+$ and $m_{\times}$, and the uncertainties on this calibration are included in the corresponding covariance matrix. This is similar to the method adopted when multiplying by $\mu$ for the angle-only method, as discussed in Section \ref{sec:likelihood}. For completeness, we have included fits for multiplicative and additive biases, and the corresponding $Q_{\mathrm{c}}$-value for the calibrated {\tt IM3SHAPE} shear estimates in Table \ref{table:m_and_c}.

Finally, we look at angle-only shear estimates calculated using the uncalibrated {\tt IM3SHAPE} ellipticities. These are given in Table \ref{table:m_and_c} as ``{\tt IM3SHAPE}: angle-only''. We see that the multiplicative bias has been reduced when using the angle-only approach with the uncalibrated ellipticities. However, the {\tt IM3SHAPE} angle-only method does require an estimate of $\mu$ and, due to differences between the errors on the position angle estimates, the $F_1$ function for {\tt IM3SHAPE} angle-only is different to that when using RPC. For the {\tt IM3SHAPE} angle-only analysis, we estimate $\mu$ directly from the simulated data set using the known input shear values, for simplicity. The value is found to be $\hat{\mu}=0.508\pm0.009$. We also see that the additive bias is smaller for the {\tt IM3SHAPE} angle-only shears than for those using the RPC method; however, we assume the correct PSF profiles when using {\tt IM3SHAPE}, whereas there is noise on the PSF images when using RPC.

\section{Shear power spectra}
\label{sec:ps}

We infer cosmology from the shear maps constructed in the previous section by estimating the shear power spectrum for each $z$-bin combination. The power spectra are reconstructed using the {\tt HEALPIX} subroutine {\tt anafast}. This routine provides pseudo-$C_l$ estimates, $\hat{C}^{ij}_l$ , with
\begin{equation}\label{eq:cl_g}
  \hat{C}^{ij}_l=\frac{1}{2l+1}\sum_{m=-l}^{+l}\tilde{g}^i_{lm}\left(\tilde{g}^j_{lm}\right)^*,
\end{equation}
where $\tilde{g}_{lm}$ denotes the spherical harmonic coefficients of the shear maps and the postscripts denote the $z$-bins being correlated.

Assuming the shear estimator in equation (\ref{eq:linest_maps}), the pseudo-$C_l$ estimates in equation (\ref{eq:cl_g}) can be written in terms of the spherical harmonic coefficients of the cosine and sine maps, $\tilde{C}_{lm}$ and $\tilde{S}_{lm}$ respectively, as
\begin{equation}\label{eq:cl_cossin}
  \hat{C}^{ij}_l=\frac{\hat{\mu}^2}{2l+1}\sum_{m=-l}^{+l}\tilde{T}^i_{lm}\left(\tilde{T}^j_{lm}\right)^*,
\end{equation}
where $\tilde{T}_{lm}=\tilde{C}_{lm}+i\tilde{S}_{lm}$\footnote{As as aside, in most weak lensing studies to date, the shear 2PCFs have been used to infer cosmology. From the maps of shear estimates given by equation (\ref{eq:linest_maps}), the 2PCFs can be written in terms of the sine and cosine maps, where
\begin{equation}
\xi_{\pm}^{ij}=\mu^2\left<C_i\left(\theta'\right)C_j\left(\theta-\theta'\right)\pm S_i\left(\theta'\right)S_j\left(\theta-\theta'\right)\right>.\nonumber
\end{equation}}. This form of the pseudo-$C_l$ estimator is useful when calculating the covariance matrix of the $C_l$ estimates, as discussed in the next section.

If one assumes full-sky and unbiased shear estimates, the expectation value of $\hat{C}^{ij}_l$ is
\begin{equation}
  \left<\hat{C}_l^{ij}\right>=C_{l}^{ij}+N_{l}^{ij},
\end{equation}
where $C_{l}^{ij}$ are the true shear power spectra and $N_l^{ij}$ are the noise power spectra, which we will now approximate analytically.

A field of shear estimates can be written in terms of the true shear field and a field of estimate errors, $\delta_{g}(\bm{\theta})$,
\begin{equation}
\hat{g}\left(\bm{\theta}\right)=g\left(\bm{\theta}\right)+\delta_{g}\left(\bm{\theta}\right).
\end{equation}
The noise power spectrum of this field is then the power spectrum of $\delta_{g}$. Assuming the shear estimator in equation (\ref{eq:linest}), the dispersion in $\delta_g$ is given by
\begin{equation}\label{eq:disp_err}
\sigma^2_{g_{1,2}}=\frac{\mu^2}{2N}\left[1-F_1^2\pm\left(F_2-F_1^2\right)\cos\left(4\alpha_0\right)\right],
\end{equation}
where $N$ is the number of galaxies in the sample used to estimate the shear. As with the function $F_1$, $F_2$ follows the definition in equation (\ref{eq:Fn_cosdelta}). If $|g|$ is much less than dispersion in galaxy ellipticities, this becomes
\begin{equation}\label{eq:error_g}
\sigma^2_{g_{1,2}}=\sigma^2_g\approx\frac{\mu^2}{2N},
\end{equation}
where for the ideal case of zero measurement errors, the factor $\mu^2/2$ depends on the intrinsic shape distribution and is analogous to shape noise for ellipticity based shear estimators. In the more general case, the value of $\mu$ depends on both the intrinsic shape distribution and the distribution of measurement errors. The noise power spectra can then be approximated as
\begin{equation}\label{eq:nls}
N_l^{ij}\approx\delta_{ij}\frac{\mu^2}{2n_i},
\end{equation}
where $n_i$ is the number of sources per steradian in $z$-bin $i$ and $\delta_{ij}$ is the Kronecker delta function.

This approximate form for $N^{ij}_l$ relies on the assumption that we can accurately estimate the shear using equation (\ref{eq:linest}). As discussed in the previous section, this estimator appears to work well with this simulated data set. However, in general it may be the case that the $F_1$ function is not accurately approximated by a first-order expansion, and so $\sigma_g$ will depend on the shear. This is because it is easier to estimate the position angles of objects with large ellipticities. If the shear is large enough to significantly change the observed shapes of a considerable number of objects, the dispersion of the errors on the position angles, and so on the shear itself, will be changed. Note, this could also be true for small shears if the sources have typically low intrinsic ellipticities. An investigation into this is left for future work, but it should be pointed out that this would not be a problem for 2PCFs, since they are insensitive to this kind of noise bias.

For our simulation, we consider a pixelated cut-sky observation, and so the theoretical shear power spectra for this region of sky becomes
\begin{equation}
\tilde{C}_l^{ij}=\sum_{l'}\mathbf{M}_{ll'}W_{l'}C_{l'}^{ij},
\end{equation}
where $C_l^{ij}$ are the input power spectra to the simulation; $W_l$ is the pixel window function; and $\mathbf{M}_{ll'}$ is the mask mixing matrix (\citealt{peebles73,wandelt01,hivon02}), which describes how power is mixed between multipoles and between E- and B-modes in observed power spectra when we consider a cut sky. Since the pixelated maps are created using {\tt HEALPIX}, we adopt the spin-2 (polarisation) {\tt HEALPIX} pixel window function for $N_{\mathrm{side}}=1024$ for the simulation.

The noise power spectra for the observed region is
\begin{equation}
\tilde{N}_l^{ij}=\sum_{l'}\mathbf{M}_{ll'}N_{l'}^{ij},
\end{equation}
where $N_l^{ij}$ is given in equation (\ref{eq:nls}).

We bin our $C_l$ estimates into 10 log-spaced multipole bins on the range $30\leq l\leq2048$. The lower limit of this range is chosen to match the lower limit of multipoles for which a Gaussian likelihood was assumed in \cite{aghanim16}. The binned values considered are
\begin{equation}\label{eq:binned_ests}
\hat{\mathcal{C}}_b^{ij}=\sum_l^{l\in b}P_{bl}\left(\hat{C}_l^{ij}-\tilde{N}_l^{ij}\right),
\end{equation}
where $b$ labels the multipole bin and $P_{bl}=l(l+1)/2\pi$. The corresponding binned theory $C_l$s are
\begin{equation}\label{eq:binned_theory}
\mathcal{C}_b^{ij}=\sum_l^{l\in b}P_{bl}\sum_{l'}\mathbf{M}_{ll'}W_{l'}C_{l'}^{ij}.
\end{equation}

The mixing matrix is calculated using the prescription presented in \cite{brown05}. For this paper, we focus on E-mode power spectra only, but we have confirmed that the estimated B-mode signals are consistent with zero in all cases.

\begin{figure*}
\begin{minipage}{6in}
\centering
\includegraphics{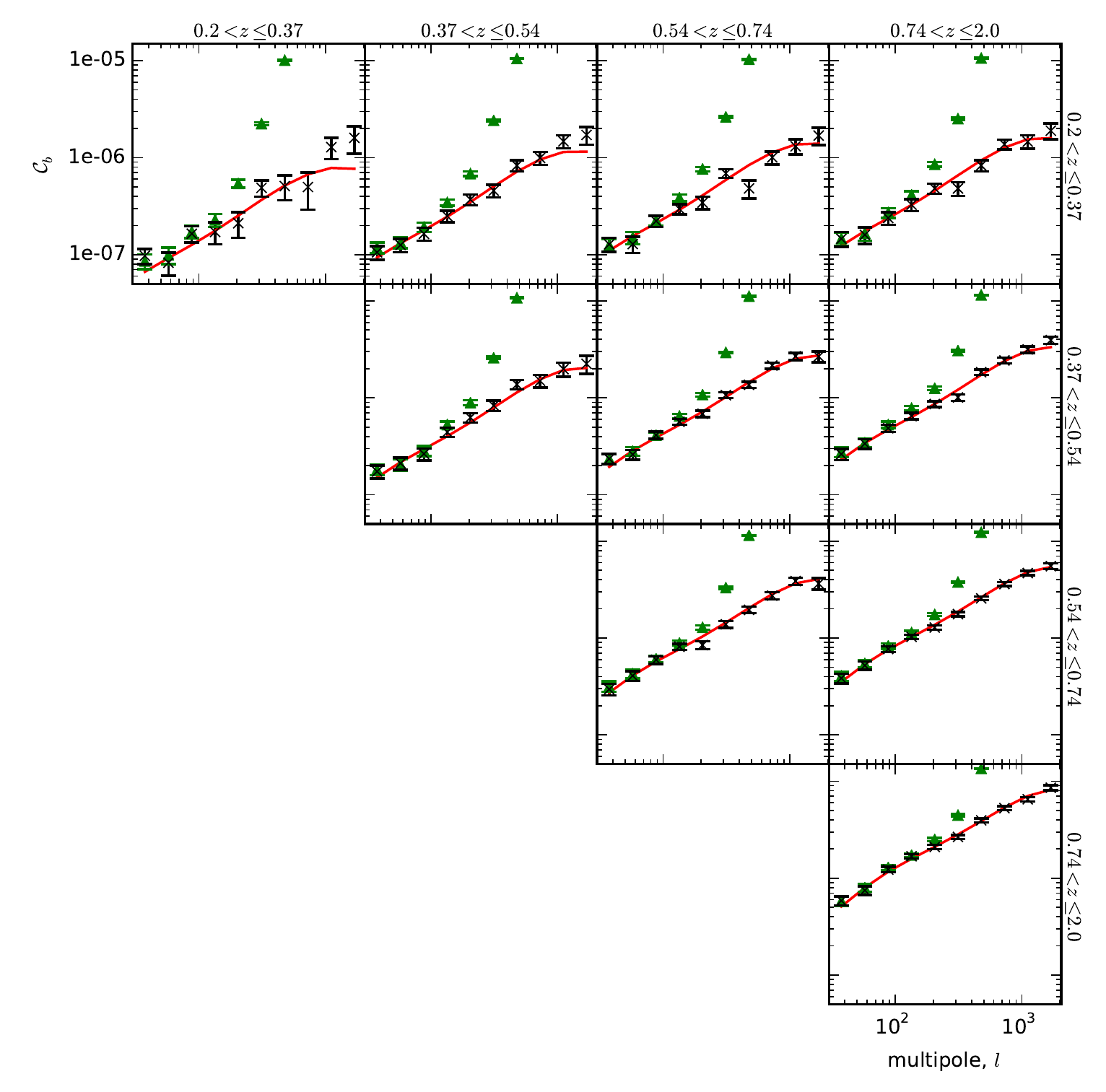}
\caption{The estimated $\mathcal{C}_b$s for each $z$-bin combination when using RPC to recover the shears (black crosses). The red curves show the theory $\mathcal{C}_b$s calculated using the input theory $C_l$s. Each $\mathcal{C}_b$ is plotted at the mean multipole value of the corresponding multipole bin. As a comparison, we also show as the green triangles the $\mathcal{C}_b$s for the case where we estimate the shears using the angle-only method without the convolution step and so without correcting for PSF anisotropy.}
\label{fig:cl_est}
\end{minipage}
\end{figure*}

The binned power spectra (equation (\ref{eq:binned_ests})) recovered from the shears estimated with the RPC method (``With RPC'' in Table \ref{table:m_and_c}) are shown in Fig. \ref{fig:cl_est} for each $z$-bin combination. The error bars, $\sigma_{\mathcal{C}_b}$, are calculated from the diagonal elements of the blocks in the covariance matrix discussed in the next section. The theory curve is given by equation (\ref{eq:binned_theory}). We see that the estimated power spectra are well described by the theory in all cases. In Fig. \ref{fig:cl_est}, we also show power spectra for the angle-only shear estimates where we estimate the position angles without using RPC (``Without RPC'' in Table \ref{table:m_and_c}). In this case, we clearly see that the power spectra are inconsistent with the theory curves.

\begin{figure*}
\begin{minipage}{6in}
\centering
\includegraphics{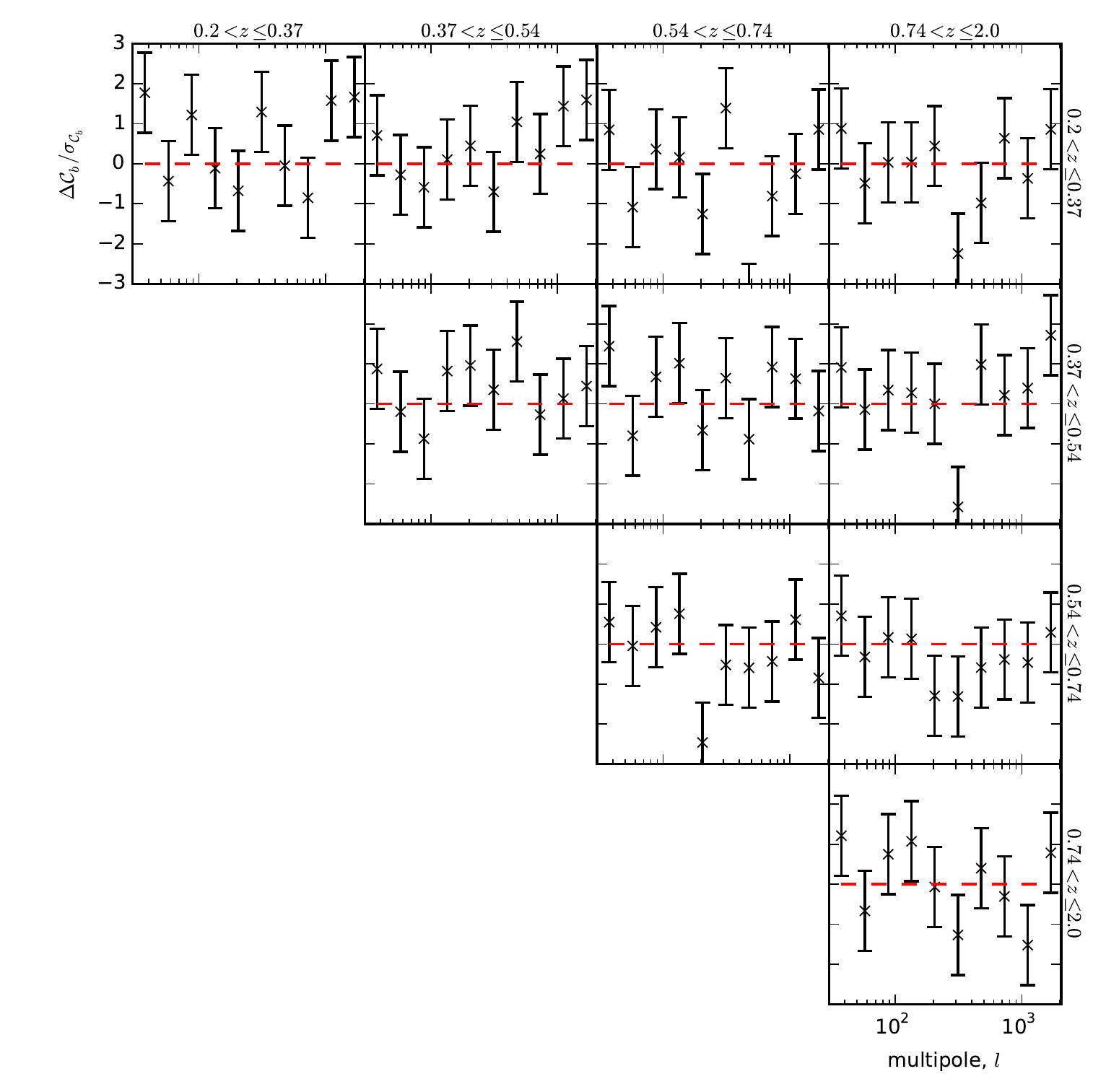}
\caption{The difference between the binned power spectra using RPC and the binned theory spectra as a fraction of the errors ($\Delta \mathcal{C}_b/\sigma_{\mathcal{C}_b}$). The red-dashed lines indicate zero difference. We see that all but one of the estimated $\mathcal{C}_b$s are within $3\sigma_{\mathcal{C}_b}$ of the input values.}
\label{fig:cl_est_bias}
\end{minipage}
\end{figure*}

As a further test for biases when using RPC, in Fig. \ref{fig:cl_est_bias} we show $\Delta \mathcal{C}_b/\sigma_{\mathcal{C}_b}$, where $\Delta \mathcal{C}_b=\hat{\mathcal{C}}_b-\mathcal{C}_b$. From this we see that all but one of the estimates lie within 3$\sigma_{\mathcal{C}_b}$ of the theory values. It should be pointed out that there will be some correlation in the errors for neighbouring bins due to the effects of the mask. This effect is accounted for in the covariance matrix when fitting for cosmology, but the effect is found to be small, as can be seen in the correlation matrix shown in Fig. \ref{fig:corr_matrix}.

To compare cosmological constraints made using angle-only shear estimates with constraints made using {\tt IM3SHAPE}, we first calibrate the {\tt IM3SHAPE} shear estimates for the multiplicative biases found in the row labelled ``{\tt IM3SHAPE}: uncalibrated'' in Table \ref{table:m_and_c}. The calibrated shears are given by
\begin{equation}
\hat{g}^{\mathrm{cal}}=\frac{\hat{g}}{\left(m_{+}+m_{\times}\right)/2}.
\end{equation}
The multiplicative and additive biases, and the $Q_{\mathrm{c}}$ value for the calibrated shears are given in the row labelled ``{\tt IM3SHAPE}: calibrated'' in Table \ref{table:m_and_c}. We calculate binned $C_l$s using these shears.

Finally, we calculate binned $C_l$s using the {\tt IM3SHAPE} angle-only shear estimates.

We do not present the power spectra recovered using {\tt IM3SHAPE} here; however, we have confirmed that they are consistent with theory using the methods discussed above.

\section{Likelihood and covariance matrix}
\label{sec:likelihood}

The binned power spectra from the previous section (shown as the black crosses in Fig. \ref{fig:cl_est}) are used to construct posteriors for the cosmological parameters $\Omega_{\mathrm{m}}$, $\sigma_8$, and $w$. To do this, we use the {\tt MULTINEST} sampler in {\tt COSMOSIS} to sample the parameters, and from the samples we calculate model $C_l$s. The model $C_l$s are binned according to equation (\ref{eq:binned_theory}). We define a data vector $\bm{\mathcal{C}}$, which contains all the estimated $\mathcal{C}^{ij}_{b}$, and a model vector $\bm{\mathcal{C}}^{\mathrm{model}}$, which contains the corresponding binned model $C_l$s. We assume the Gaussian likelihood
\begin{equation}
\mathcal{L}\propto\exp\left(-\frac{1}{2}\bm{d}^{\mathrm{T}}\mathbf{C}^{-1}\bm{d}\right),
\end{equation}
where $\bm{d}=\bm{\mathcal{C}}-\bm{\mathcal{C}}^{\mathrm{model}}$ and $\mathbf{C}$ is the covariance matrix. The vector $\bm{d}$ contains $N_{\mathrm{d}}=100$ elements. This is the number of multipole bins multiplied by the number of $z$-bin combinations. As discussed in the previous section, we focus on multipoles of $l\geq30$, which is the lower limit of multipoles for which a Gaussian likelihood is assumed by \cite{aghanim16}.

At low multipoles, the likelihood of pseudo-$C_l$ estimates is not well described by a Gaussian distribution; however, the Gaussian approximation increases in accuracy at higher multipoles \citep{sun13}, and we find no evidence of a bias in our parameter constraints (see the following section). For Stage IV weak lensing analyses based on second-order summary statistics, it may be necessary to use a likelihood based on a more exact form \citep{sellentin17}. An exact likelihood for pseudo-$C_l$ estimates for the general case of an arbitrary mask is derived in \cite{upham19}.

To estimate the covariance matrix, we begin by assuming that errors on the shear estimates within each pixel are Gaussian distributed. We then produce $N_{\mathrm{s}}=10^3$ maps of signal + noise Gaussian random simulated shear estimates for each set of $z$-bins using the same observed area and galaxy numbers as in Section \ref{sec:cos_sim}. The power spectra are calculated from these shear maps, and the covariance matrix is estimated from the power spectra. Here, we discuss the details of this method.

In order for errors on the shear estimates to be Gaussian distributed, errors on the means of the cosines and sines within each pixel must also be Gaussian distributed. We have verified this to be true for means of the trigonometric functions for which there are $\gtrsim$10 sources within a pixel: there are 50 sources per pixel in our simulation.

If we assume that we measure the position angles from a set of measured ellipticities\footnote{This assumption is made to make it easier to simulate errors on the position angles. Since the ratio $\epsilon_2/\epsilon_1$ is equal to $\tan(2\alpha)$ and the means of the cosines and sines are approximately Gaussian distributed, simulating distributions of ellipticities should be equivalent to simulating distributions of cosines and sines provided the correct ellipticity dispersions are used.} and that the errors and intrinsic ellipticities are Gaussian distributed, equation (\ref{eq:k_sigs})  gives us a relationship between $\mu$ and the corresponding 1D dispersion of the ellipticities, $\sigma_{\mathrm{RMS}}=\sqrt{\sigma_{\epsilon}^2+\sigma_{\mathrm{err}}^2}$, where $\sigma_{\mathrm{err}}$ is the error on the ellipticity measurements. Since we have already estimated $\mu$ from the constant shear simulation discussed in the previous section, we use
this value to calculate $\sigma_{\mathrm{RMS}}$.

For these simulations, we create {\tt HEALPIX} maps of Gaussian random shear fields for each $z$-bin with the same observed area as above and assuming the same input cosmology. This is done independently for each of the $10^3$ realisations. For each pixel, we produce 50 observed galaxy shapes, each with an independent random Gaussian component, $\delta_{\mathrm{RMS}}$, with zero mean and 1D dispersion $\sigma_{\mathrm{RMS}}$. The observed ellipticities are then
\begin{equation}
\epsilon^{(k,n)}=g^{(k)}+\delta_{\mathrm{RMS}}^{(k,n)},
\end{equation}
where the postscript $k$ labels the pixel and $n$ is one of the 50 galaxies within the pixel. From these shapes, we calculate the corresponding position angles and then construct mean cosine and sine maps ($C(\bm{\theta})$ and $S(\bm{\theta})$). This provides us with $10^3$ maps of mean cosines and $10^3$ maps of mean sines for each $z$-bin.

For each realisation, we calculate the power spectra as
\begin{equation}\label{eq:cl_cossin_cov}
  \hat{C}^{ij}_l=\frac{\left(\hat{\mu}+\delta_{\mu}\right)^2}{2l+1}\sum_{m=-l}^{+l}\tilde{T}^i_{lm}\left(\tilde{T}^j_{lm}\right)^*,
\end{equation}
where $\hat{\mu}$ is the estimated $\mu$ value given in the previous section ($\hat{\mu}=0.574$), and $\delta_\mu$ is a Gaussian random number with zero mean and with the dispersion quoted for $\hat{\mu}$ in the previous section. The value of $\delta_\mu$ is kept the same for all the $\hat{C}_l$s in a given realisation, but a different value is generated for each realisation. This is done to propagate the uncertainty on $\mu$ into the covariance matrix.

Each of the estimated power spectra are binned according to equation (\ref{eq:binned_ests}), and we use the $10^3$ sets of binned $C_l$s to estimate the covariance matrix, $\hat{\mathbf{C}}$.

\begin{figure}
\centering
\includegraphics{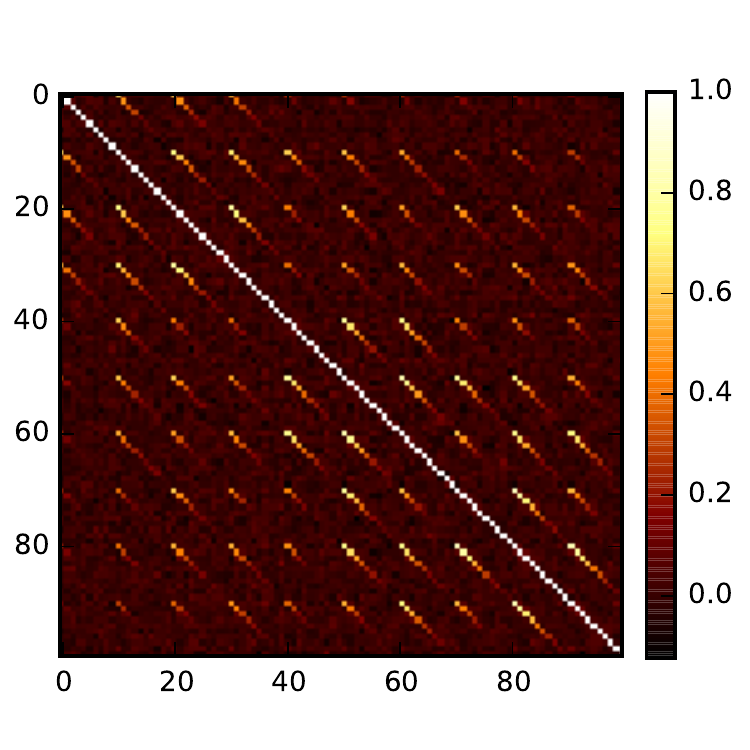}
\caption{The correlation matrix ($C_{ij}/\sqrt{C_{ii}C_{jj}}$) for the RPC binned power spectra estimated from $10^3$ Monte-Carlo simulations. The block ordering corresponds to the $z$-bin combinations 11, 12, 13, 14, 22, 23, 24, 33, 34 and 44. Each block contains the covariance of the ten bandpowers discussed in the main text. As can be seen, the blocks are dominated by the diagonal elements, indicating that neighbouring multipole bins are approximately uncorrelated.}
\label{fig:corr_matrix}
\end{figure}

In Fig. \ref{fig:corr_matrix}, we show the correlation matrix, which is the covariance matrix normalised by the diagonal components. From this, we see that there appears to be very little correlation between neighbouring multipole bins.

Given an estimate of the covariance matrix, $\hat{\mathbf{C}}$, an unbiased estimate of the inverse covariance matrix is (\citealt{anderson03, hartlap07})
\begin{equation}
\hat{\Psi}=\frac{N_{\mathrm{s}}-N_{\mathrm{d}}-2}{N_{\mathrm{s}}-1}\hat{\mathbf{C}},
\end{equation}
and so the likelihood we use to infer cosmology is
\begin{equation}\label{eq:like_use}
\mathcal{L}\propto\exp\left(-\frac{1}{2}\bm{d}^{\mathrm{T}}\hat{\mathbf{\Psi}}^{-1}\bm{d}\right).
\end{equation}

The above procedure is repeated for both the calibrated {\tt IM3SHAPE} $C_l$ estimates and the {\tt IM3SHAPE} angle-only $C_l$ estimates to provide us with their corresponding covariance matrices and likelihoods. Note that, for the case of the {\tt IM3SHAPE} $C_l$s, errors on the multiplicative bias calibration are propagated using the same approach as that used to propagate errors on $\mu$ for the angle-only cases.

\section{Results}
\label{sec:results}
Here, we present the main results of the paper. Posteriors are constructed for the parameters $\Omega_{\mathrm{m}}$, $\sigma_8$, and $w$ using the binned $C_l$s recovered from the RPC angle-only shear estimates, as discussed in Section \ref{sec:ps}, and the likelihood given in equation (\ref{eq:like_use}). We fit for a $w$CDM model and fix all the other parameters to their fiducial values. The posteriors are plotted using {\tt GETDIST}\footnote{\url{https://github.com/cmbant/getdist}}.

\begin{figure*}
\begin{minipage}{6in}
\centering
\includegraphics{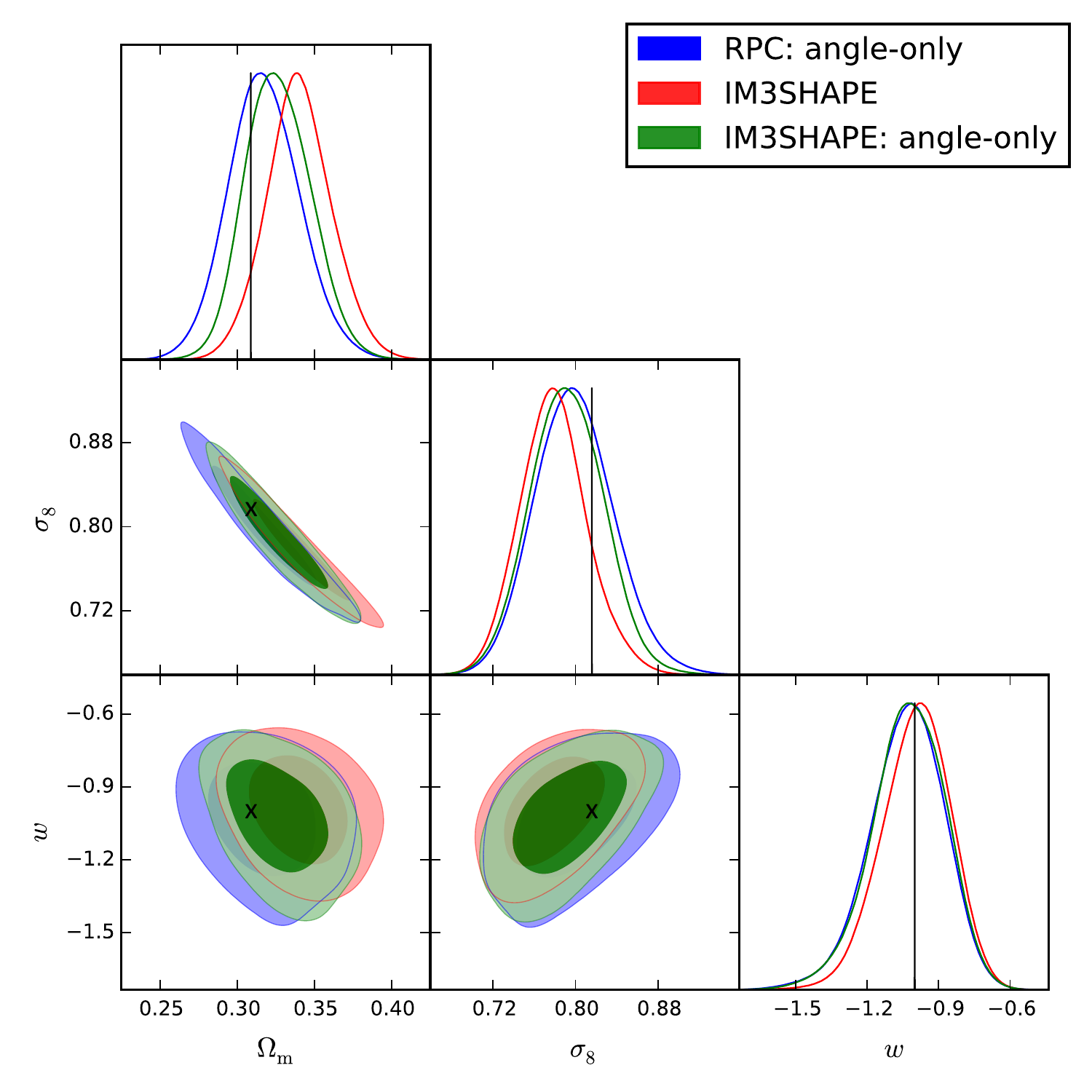}
\caption{The 1D and 2D marginalised posteriors for the RPC, {\tt IM3SHAPE}, and {\tt IM3SHAPE} angle-only analyses. The vertical black lines and the black crosses indicate the input values of the parameters.}
\label{fig:cos_posts}
\end{minipage}
\end{figure*}

\begin{table}
\centering
\begin{tabular}{|c|c|c|c|}
\hline
Parameter & RPC: angle-only & {\tt IM3SHAPE} & {\tt IM3SHAPE}: angle-only \\ [0.5ex]
\hline
$\Omega_m$ &  $0.318\pm0.023$ & $0.338^{+0.021}_{-0.020}$ & $0.327\pm0.021$ \\ [2ex]
$\sigma_8$ &  $0.799\pm0.038$ & $0.779^{+0.031}_{-0.033}$ & $0.729\pm0.035$ \\  [2ex]
$w$ &  $-1.03^{+0.16}_{-0.15}$ & $-0.99^{+0.14}_{-0.15}$ & $-1.03\pm0.15$ \\ [1ex]
\hline
\end{tabular}
\caption{Marginalised mean values and 95\% confidence limits for the fitted parameters.}
\label{table:cos_param}
\end{table}

In addition to the posteriors constructed from the RPC angle-only $C_l$ estimates, we produce posteriors for the $C_l$s recovered using the calibrated {\tt IM3SHAPE} shear estimates and the {\tt IM3SHAPE} angle-only shear estimates. The 1D and 2D marginalised posteriors are shown in Fig. \ref{fig:cos_posts}, with the means and 68\% confidence limits of the parameters given in Table \ref{table:cos_param}. From Fig. \ref{fig:cos_posts}, we see that the constraining power of the angle-only methods are similar to that when using {\tt IM3SHAPE} with full-ellipticity information. There does appear to be a small increase in the area of the contours when going from {\tt IM3SHAPE} (both full-ellipticity and angle-only) to the RPC angle-only method. This may be partially due to assuming the correct PSF profiles when using {\tt IM3SHAPE} and including noise on the PSF images when using RPC.

When going from full-ellipticity information to angle-only information using {\tt IM3SHAPE}, we find no clear increase in the uncertainty on $\Omega_{\mathrm{m}}$, but we do find an increase in the standard deviation of the 1D marginalised likelihoods for $\sigma_8$ and $w$ of $\sim$9\% and $\sim$8\% respectively. We also expect there to be an increase in the uncertainty on $\Omega_{\mathrm{m}}$. Due to the computation time required to perform these tests, we only ran the simulation once, but the shape of the posterior is expected to change for different realisations; therefore, there is some noise associated with the derived uncertainties on the parameters. If we were to repeat this test for multiple realisations, we expect we would also identify an increase in the uncertainty on $\Omega_{\mathrm{m}}$.

In all cases, the input values of the parameters lie within the 95\% confidence contours, and so there is no clear indication that
systematics pose a problem.

To investigate the level of bias in the parameters when using these methods, one would need to run many simulations. This level of detail is beyond the scope of this paper. Here, we have demonstrated that there is no evidence that angle-only methods are dominated by systematics when considering a Stage III weak lensing simulation.

When estimating the value of $\mu$ for the RPC method, the calibration simulation was identical to the simulation used for the observed data set, only with a fixed shear modulus and a different set of random seeds. Therefore, this simulation contained the correct PSF profiles, noise levels, galaxy sizes, and intrinsic ellipticity distribution. In reality, however, this information would also have associated errors. This would increase the uncertainty on $\mu$ and therefore the size of the confidence intervals. Future work will look at how sensitive angle-only weak lensing is to these kinds of errors, but we should point out that these kinds of uncertainties would also be present in the simulations used to calibrate shear estimates that use full-ellipticity information.

\section{Conclusion}
\label{sec:conc}
We have demonstrated that it is possible to constrain cosmology using only the statistics of source galaxy position angles in the presence of a variable elliptical PSF. The goal of this paper is to build on previous work on angle-only shear estimators by extending the approach to deal with variable shear fields and removing the necessity to downweight the contribution from sources aligned with the PSF. The application of angle-only shear estimators to the cosmological simulation in Section \ref{sec:cos_sim} successfully demonstrates the methods on variable shear fields, while also allowing us to compare their constraining powers with an estimator using full-ellipticity information. We have also shown that the effects of PSF ellipticity can be corrected for by convolving an observed galaxy image with an image of the PSF rotated by $90^{\circ}$ (the rotated PSF convolution (RPC) method). This approach successfully removes the need to downweight sources aligned with the PSF.

For the simulation discussed in Section \ref{sec:cos_sim}, we found that we could accurately recover estimates of the shear to within an overall scale factor using only position angles measured with the RPC method. The scale factor was successfully estimated using a cosmology independent calibration simulation with a fixed shear magnitude. In general, it may be that a more precise estimate of the full $F_1$ function is required, but one should be able to measure this by simply running calibration simulations for a selection of shear magnitudes.

It is common practice for shape measurement techniques to use simulations to calibrate for multiplicative and additive biases.
The method that we propose for estimating the $F_1$ function uses the same approach to calibrate the moduli of the shear estimates. For out tests we assumed an ideal set of calibration simulations, and the impact of uncertainties in the details of the simulations will be investigated in future work.

As a comparison with the RPC approach, we also estimated the shears using {\tt IM3SHAPE}. This was done using both full-ellipticity information and angle-only information. Here, we found that multiplicative biases were smaller when using {\tt IM3SHAPE} position angles than when using uncalibrated shapes. There was also a smaller additive bias for the {\tt IM3SHAPE} angle-only shear estimates than for those using the RPC method; however, noise was added to the PSF images when using the RPC method, but the correct PSF profiles were assumed for {\tt IM3SHAPE}.

To measure the position angles of the sources when using RPC, we used quadrupole moments. With this approach, we found that the angle-only estimator provided shear estimates (to within an overall scale factor) $\sim$10 faster than {\tt IM3SHAPE}, with the additional advantage that moments are model independent, and so are free from model bias.

Using the angle-only method, we demonstrated that we can potentially recover accurate estimates of shear power spectra and reliable cosmological constraints from Stage III weak lensing experiments by considering a Dark Energy Survey-like set of simulated observations. The constraining power using the angle-only method was shown to be similar to that found when using full-ellipticity information from {\tt IM3SHAPE}. We identified small increases in the uncertainties on $\sigma_8$ and $w$ of 9\% and 8\% respectively when going from {\tt IM3SHAPE} using full-ellipticity information to {\tt IM3SHAPE} using only the position angles. No clear increase in uncertainty was observed for $\Omega_{\mathrm{m}}$, but this is likely due to the statistical limitations of using a single realisation of the data.

This work indicates that angle-only shear estimation is a promising tool for weak lensing studies, and an application of the methods in this paper should meet the requirements for Stage III surveys. Future Stage IV weak lensing experiments, such as with Euclid and using the LSST, have more stringent constraints on multiplicative and additive biases. The simulation presented in this paper is not sufficient to demonstrate that residual biases in the angle-only method are below these thresholds. In future work, we intend to test the angle-only method using Stage IV-like simulations that include more realistic galaxy and PSF morphologies, and more realistic noise models. If residual biases are shown to be above the required thresholds, it may be possible to mitigate them by upsampling the PSF and galaxy images when using RPC.

\section*{Acknowledgments}
LW is grateful to Dan Thomas, Benjamin Joachimi, and Michael Brown for useful discussions. LW is supported by a UK Space Agency grant.

\bibliographystyle{mnras} \bibliography{ms}

\appendix

\section{The $F_1$ function}
\label{ap:F1}
Following the definition of the $F_n$ functions in equation (\ref{eq:Fn_cosdelta}), we present the expression for $F_1$ given in \cite{whittaker14}, which assumes negligible measurement errors, and we also extend the work to include information about measurement errors.

If we ignore errors for the moment, the $F_1$ function can be expressed in terms of the modulus of the shear, $|g|$, and the intrinsic ellipticity distribution, $f(|\epsilon^{\mathrm{int}}|)$, and is (\cite{whittaker14})
\begin{align}\label{eq:general_F}
F_1\left(\left|g\right|\right)=\, &\frac{1}{\pi}\int_0^{\left|\bm{\epsilon}_{\mathrm{max}}^{\mathrm{int}}\right|}\int_{-\frac{\pi}{2}}^{\frac{\pi}{2}}\mathrm{d}\alpha^{\mathrm{int}}\mathrm{d}\left|\epsilon^{\mathrm{int}}\right|f\left(\left|\epsilon^{\mathrm{int}}\right|\right)\nonumber\\
&\times h_1\left(\left|g\right|,\left|\epsilon^{\mathrm{int}}\right|,\alpha^{\mathrm{int}}\right).
\end{align}
Here $\alpha^{\rm int}$ is the intrinsic position angle, and the function $h_1\left(\left|g\right|,\left|\epsilon^{\mathrm{int}}\right|,\alpha^{\mathrm{int}}\right)$ is found to be
\begin{equation}\label{eq:g_function}
h_1\left(\left|g\right|,\left|\epsilon^{\mathrm{int}}\right|,\alpha^{\mathrm{int}}\right)=\frac{\epsilon_1'}{\sqrt{\epsilon_1'^2+\epsilon_2'^2}},
\end{equation}
with
\begin{align}\label{eq:e'}
\epsilon_1'=&\left|g\right|\left(1+\left|\epsilon^{\mathrm{int}}\right|^2\right)+\left(1+\left|g\right|^2\right)\left|\epsilon^{\mathrm{int}}\right|\cos\left(2\alpha^{\mathrm{int}}\right),\nonumber\\
\epsilon_2'=&\left(1-\left|g\right|^2\right)\left|\epsilon^{\mathrm{int}}\right|\sin\left(2\alpha^{\mathrm{int}}\right).
\end{align}

In reality, of course, we also need to consider measurement errors. To do this, we can assume that we recover the position angles from a set of ellipticity measurements. This is equivalent to estimating the position angles from a set of Stokes parameters (as is done in Section \ref{subsec:demo}), since $v/u=\epsilon^{\mathrm{obs}}_2/\epsilon^{\mathrm{obs}}_1$. In this case, the $F_1$ function is
\begin{align}\label{eq:F1_errdist}
F_1\left(\left|g\right|\right)=&\frac{1}{\pi}\int_0^{\left|\epsilon_{\mathrm{max}}^{\mathrm{int}}\right|}\int_{-\frac{\pi}{2}}^{\frac{\pi}{2}}\int_{\left|\epsilon^{\mathrm{obs}}\right|\leq1}\mathrm{d}^2\bm{\epsilon}^{\mathrm{obs}}\mathrm{d}\alpha^{\mathrm{int}}\mathrm{d}\left|\epsilon^{\mathrm{int}}\right|\nonumber\\
&\times\cos\left[2\delta\left(\bm{\epsilon}^{\mathrm{obs}}\right)\right]f_{\epsilon}\left(\bm{\epsilon}^{\mathrm{obs}}|\epsilon^{\mathrm{lens}},\bm{D}\right)f\left(\left|\epsilon^{\mathrm{int}}\right|\right),
\end{align}
where $\epsilon^{\mathrm{lens}}$ is a function of the shear (equation (\ref{eq:lensed_ellip})). The offset $\delta$ can be written as a function of $\epsilon^{\mathrm{obs}}$ and $|g|$ if one assumes that errors are independent of shear orientation. The function $f_{\epsilon}$ is the distribution of measured ellipticities given the true lensed ellipticity and a vector $\bm{D}$ containing other information related to the error distribution (such as galaxy sizes, signal-to-noise ratios, etc). For a given galaxy, the distribution of $\epsilon^{\mathrm{obs}}$ is expected to follow the Marsaglia-Tin distribution (\citealt{marsaglia65, tin65, viola14}) if one uses moments to estimate ellipticities from noisy images. If the vector $\bm{D}$ is identical for each galaxy used to estimate the shear, then $f_{\epsilon}$ is the Marsaglia-Tin distribution; however, in practice, there will also be a distribution of $\bm{D}$ and so $f_{\epsilon}$ is a convolution of the Marsaglia-Tin distribution with the distribution of $\bm{D}$.

To gain some insight into the nature of $F_1$, let us assume that we measure the position angles from a set of ellipticity measurements and that the observed ellipticity of an object can be simplified as
\begin{equation}\label{eq:simp_obs}
\epsilon^{\mathrm{obs}}=g+\epsilon^{\mathrm{int}}+\delta_{\epsilon},
\end{equation}
where $\delta_{\epsilon}$ is an error on the ellipticity. For equation (\ref{eq:simp_obs}) to be accurate, we require $g,\epsilon^{\mathrm{int}}\ll1$.

If both $\epsilon^{\mathrm{int}}$ and $\delta_{\epsilon}$ are Gaussian distributed with 1D dispersions $\sigma_{\epsilon}$ and $\sigma_{\mathrm{err}}$ respectively, the $F_1$ function can be Taylor expanded to first order in $|g|/\sqrt{\sigma_{\epsilon}^2+\sigma_{\mathrm{err}}^2}$,
\begin{equation}
F_1\left(\left|g\right|\right)\approx\frac{\left|g\right|}{\mu},
\end{equation}
where
\begin{equation}\label{eq:k_sigs}
\mu=\sqrt{\frac{8\left(\sigma_{\epsilon}^2+\sigma_{\mathrm{err}}^2\right)}{\pi}}.
\end{equation}

The $F_2$ function is useful when discussing errors on the shear estimates (see equation (\ref{eq:disp_err})).With the same assumptions as above, the $F_2$ function can be written to leading order in $|g|/\sqrt{\sigma_{\epsilon}^2+\sigma_{\mathrm{err}}^2}$ as
\begin{equation}
F_2\left(\left|g\right|\right)=\frac{\left|g\right|^2}{4\left(\sigma_{\epsilon}^2+\sigma_{\mathrm{err}}^2\right)}.
\end{equation}
It should be emphasised that these results for $F_1$ and $F_2$ are only true for the assumptions that $g,\epsilon^{\mathrm{int}}\ll1$, and $\epsilon^{\mathrm{int}}$ and $\delta_{\epsilon}$ are Gaussian distributed.

If errors on the position angle measurements can be considered negligible (i.e. we can adopt the $F_1$ function in equation (\ref{eq:general_F})), a general expression for $\mu$ is
\begin{equation}\label{eq:taylor_k}
\mu=\left[\int_0^{\left|\epsilon_{\mathrm{max}}^{\mathrm{int}}\right|}\mathrm{d}\left|\epsilon^{\mathrm{int}}\right|\frac{\left(1+\left|\epsilon^{\mathrm{int}}\right|^2\right)}{2\left|\epsilon^{\mathrm{int}}\right|}f\left(\left|\epsilon^{\mathrm{int}}\right|\right)\right]^{-1}.
\end{equation}

\section{Moments of image with rotated PSF convolution}
\label{ap:rot_convol}
Here we show that the rotation effects arising from anisotropic PSF convolution can be removed by further convolving the image with an image of the PSF rotated by $90^{\circ}$.

The position angle of a source galaxy can be expressed in terms of the second-order moments of the galaxy intensity profile, $I_{\mathrm{gal}}\left(x,y\right)$:
\begin{equation}
\alpha=\frac{1}{2}\tan{-1}\left(\frac{2Q_{11}}{Q_{20}-Q_{02}}\right),
\end{equation}
where the (unnormalised) second-order moments are
\begin{equation}
Q_{ij}=\int\mathrm{d}x\mathrm{d}yI_{\mathrm{gal}}\left(x,y\right)x^iy^j,
\end{equation}
and where we assume that the axes are defined such that the centroid of the image is at the origin.

Upon equivalently defining the second-order moments of the PSF as $P_{ij}$, it can be shown that the moments of the PSF-convolved galaxy image are \citep{flusser98}
\begin{equation}
\tilde{Q}_{ij}=\sum_i^k\sum_j^l{k\choose i}{l\choose j}Q_{kl}P_{k-i,l-j}.
\end{equation}
This is a general results that holds for any galaxy and PSF morphology as long as their moments do not diverge. The second moments of the convolved image are then
\begin{align}
Q^{*}_{20}=&Q_{20}+Q_{00}P_{20},\nonumber\\
Q^{*}_{02}=&Q_{02}+Q_{00}P_{02},\nonumber\\
Q^{*}_{11}=&Q_{11}+Q_{00}P_{11}.
\end{align}
The total flux of the convolved image is $Q^{*}_{00}=Q_{00}$.

If we assume that we have high-resolution noise free images of both the PSF and the PSF-convolved galaxy, we can first rotate the PSF image by $90^{\circ}$ and convolve our already PSF convolved galaxy image with the rotated PSF image, so that
\begin{align}\label{eq:moments}
Q^{*\perp}_{20}=&Q^{*}_{20}P^{\perp}_{00}+Q^{*}_{00}P^{\perp}_{20},\nonumber\\
Q^{*\perp}_{02}=&Q^{*}_{02}P^{\perp}_{00}+Q^{*}_{00}P^{\perp}_{02},\nonumber\\
Q^{*\perp}_{11}=&Q^{*}_{11}P^{\perp}_{00}+Q^{*}_{00}P^{\perp}_{11}.
\end{align}
where $P^{\perp}$ is the rotated PSF image, which has total flux $P^{\perp}_{00}$. If the PSF image is normalised, $P^{\perp}_{00}=1$.

The second-order moments of the rotated PSF image are
\begin{align}
P^{\perp}_{20}=&P^{\perp}_{00}P_{02},\nonumber\\
P^{\perp}_{02}=&P^{\perp}_{00}P_{20},\nonumber\\
P^{\perp}_{11}=&-P^{\perp}_{00}P_{11},
\end{align}
so that the moments in equation (\ref{eq:moments}) are given by
\begin{align}\label{eq:moments2}
Q^{*\perp}_{20}=&Q_{20}P^{\perp}_{00}+Q_{00}P^{\perp}_{00}\left(P_{20}+P_{02}\right),\nonumber\\
Q^{*\perp}_{02}=&Q_{02}P^{\perp}_{00}+Q_{00}P^{\perp}_{00}\left(P_{20}+P_{02}\right),\nonumber\\
Q^{*\perp}_{11}=&Q_{11}P^{\perp}_{00},
\end{align}
and therefore
\begin{align}\label{eq:moments3}
Q^{*\perp}_{20}-Q^{*\perp}_{02}=&P^{\perp}_{00}\left(Q_{20}-Q_{02}\right),\nonumber\\
Q^{*\perp}_{11}=&P^{\perp}_{00}Q_{11}.
\end{align}
Hence, the position angle of the image convolved with the rotated PSF is identical to that of the source galaxy, whilst the Stokes parameters (see equation (\ref{eq:stokes_params})) are scaled by a factor of $P^{\perp}_{00}$. This result is completely general and makes no assumptions about the models of the galaxy or PSF.

\section{Distribution of errors on position angle estimates}
\label{ap:err_posang}

Here we derive an expression for the distribution of position angle estimates, given Gaussian pixel noise on both the galaxy image and the PSF image, when using quadrupole moments with the RPC method.

For convenience, we start by defining the functions
\begin{align}\label{eq:F12}
U_{\mathrm{gal}}\left(\bm{\theta}_{ij}\right)=&\sum_{kl}\left(x_k^2-y_l^2\right)W\left(\bm{\theta}_{kl}\right)I^{*}_{\mathrm{gal}}\left(\bm{\theta}_{ij}-\bm{\theta}_{kl}\right),\nonumber\\
V_{\mathrm{gal}}\left(\bm{\theta}_{ij}\right)=&\sum_{kl}\left(2x_ky_l\right)W\left(\bm{\theta}_{kl}\right)I^{*}_{\mathrm{gal}}\left(\bm{\theta}_{ij}-\bm{\theta}_{kl}\right),\nonumber\\
U_{\mathrm{PSF}}\left(\bm{\theta}_{ij}\right)=&\sum_{kl}\left(x_k^2-y_l^2\right)W\left(\bm{\theta}_{kl}\right)I^{\perp}_{\mathrm{PSF}}\left(\bm{\theta}_{ij}-\bm{\theta}_{kl}\right),\nonumber\\
V_{\mathrm{PSF}}\left(\bm{\theta}_{ij}\right)=&\sum_{kl}\left(2x_ky_l\right)W\left(\bm{\theta}_{kl}\right)I^{\perp}_{\mathrm{PSF}}\left(\bm{\theta}_{ij}-\bm{\theta}_{kl}\right),
\end{align}
which involve convolutions with images of the PSF-convolved galaxy and the rotated PSF. These functions are defined such that the Stokes parameters of the PSF-convolved source, $I^{*}_{\mathrm{gal}}$ (without noise), are
\begin{align}
u^{*}=&\sum_{ij}U_{\mathrm{gal}}\left(\bm{\theta}_{ij}\right),\nonumber\\
v^{*}=&\sum_{ij}V_{\mathrm{gal}}\left(\bm{\theta}_{ij}\right),
\end{align}
and, if the mean of the pixel noise is zero, the expectation values of the Stokes parameters (equation (\ref{eq:stokes_params})) are
\begin{align}\label{eq:stokes_FG}
\left<\hat{u}\right>=&\sum_{ij}I^{*}_{\mathrm{gal}}\left(\bm{\theta}_{ij}\right)U_{\mathrm{PSF}}\left(\bm{\theta}_{ij}\right),\nonumber\\
\left<\hat{v}\right>=&\sum_{ij}I^{*}_{\mathrm{gal}}\left(\bm{\theta}_{ij}\right)V_{\mathrm{PSF}}\left(\bm{\theta}_{ij}\right).
\end{align}
The coordinates $\bm{\theta}_{ij}$ are assumed to have their origin at the centroid of $I^{*}_{\mathrm{gal}}$ (the PSF-convolved source). We assume that we know the centroid exactly and use a fixed-size weight function.

In the absence of errors on the centroid estimates and assuming a negligible contribution from pixelation, the expectation values of the Stokes parameters are
\begin{align}\label{eq:stokes_mom}
\left<\hat{u}\right>=P^{\perp}_{00}u_0,\nonumber\\
\left<\hat{v}\right>=P^{\perp}_{00}v_0,
\end{align}
as discussed in Appendix \ref{ap:rot_convol}, and where $u_0$ and $v_0$ are the Stokes parameters of the source, $I_{\mathrm{gal}}$ (i.e. prior to PSF convolution), and $P^{\perp}_{00}$ is the total flux of the rotated PSF image, which will be unity if the image is normalised. Therefore,
\begin{equation}
\frac{1}{2}\tan^{-1}\left(\frac{\left<\hat{v}\right>}{\left<\hat{u}\right>}\right)=\frac{1}{2}\tan^{-1}\left(\frac{v_0}{u_0}\right)=\alpha,
\end{equation}
as required.

It can be shown that, if one uses an isotropic weight function, the covariance matrix of $\hat{u}$ and $\hat{v}$ has components 
\begin{align}\label{eq:cov_comp}
\sigma_u^2\equiv&\left<\left(\hat{u}-\left<\hat{u}\right>\right)^2\right>=\sum_{ij}\left[\sigma_{\mathrm{I}}^2U_{\mathrm{PSF}}^2\left(\bm{\theta}_{ij}\right)+\sigma_{\mathrm{PSF}}^2U_{\mathrm{gal}}^2\left(\bm{\theta}_{ij}\right)\right],\nonumber\\
\sigma_v^2\equiv&\left<\left(\hat{v}-\left<\hat{v}\right>\right)^2\right>=\sum_{ij}\left[\sigma_{\mathrm{I}}^2V_{\mathrm{PSF}}^2\left(\bm{\theta}_{ij}\right)+\sigma_{\mathrm{PSF}}^2V_{\mathrm{gal}}^2\left(\bm{\theta}_{ij}\right)\right],\nonumber\\
\sigma_{uv}^2\equiv&\left<\left(\hat{u}-\left<\hat{u}\right>\right)\left(\hat{v}-\left<\hat{v}\right>\right)\right>=\nonumber\\
\sum_{ij}&\left[\sigma_{\mathrm{I}}^2U_{\mathrm{PSF}}\left(\bm{\theta}_{ij}\right)V_{\mathrm{PSF}}\left(\bm{\theta}_{ij}\right)+\sigma_{\mathrm{PSF}}^2U_{\mathrm{gal}}\left(\bm{\theta}_{ij}\right)V_{\mathrm{gal}}\left(\bm{\theta}_{ij}\right)\right],
\end{align}
where $\sigma_{\mathrm{I}}^2$ is the variance of the pixel noise in the galaxy image and $\sigma_{\mathrm{PSF}}^2$ is the variance of the pixel noise in the PSF image.

As the moments are linear functions of the (galaxy and PSF) image pixels, if the pixel noise is assumed to be Gaussian distributed, errors on the Stokes parameters are also Gaussian distributed. In this case, given the covariance matrix
\begin{equation}\label{eq:R}
\mathbf{C}=\left(
\begin{array}{cc}
\sigma_u^2 & \sigma_{uv}^2\\
\sigma_{uv}^2 & \sigma_v^2
\end{array}\right),
\end{equation}
the distribution of estimated position angles, $\hat{\alpha}$, is
\begin{equation}\label{eq:marg_chi}
f\left(\hat{\alpha}|\left<u\right>,\left<v\right>\right)=K\left[\frac{e^{-AG^2}}{A}+G\sqrt{\frac{\pi}{A}}\left(1+\mathrm{erf}\left(G\sqrt{A}\right)\right)\right],
\end{equation}
where
\begin{align}
A=&\frac{1}{2}\bm{t}^{\mathrm{T}}\mathbf{C}^{-1}\bm{t},\nonumber\\
G=&\frac{1}{2A}\bm{x}_0^{\mathrm{T}}\mathbf{C}^{-1}\bm{t},\nonumber\\
K=&\frac{1}{2\pi\sqrt{\mathrm{det}[\mathbf{C}]}}\exp\left[-\frac{1}{2}\left(\bm{x}_0^{\mathrm{T}}\mathbf{C}^{-1}\bm{x}_0-2AG^2\right)\right],
\end{align}
and we have defined
\begin{equation}\label{eq:xvec}
\bm{x}_0=\left(
\begin{array}{c}
 \left<u\right>\\
 \left<v\right>
\end{array} \right),
\end{equation}
\begin{equation}\label{eq:tvec}
\bm{t}=\left(
\begin{array}{c}
 \cos\left(2\hat{\alpha}\right)\\
 \sin\left(2\hat{\alpha}\right)
\end{array} \right).
\end{equation}

If one has an accurate distribution describing the errors on the centroid, one may write the $U$ and $V$ functions in equation (\ref{eq:F12}) as functions of the centroid estimates. Upon doing this, one may convolve the distribution in equation (\ref{eq:marg_chi}) with the centroid errors to give the full distribution of errors on $\alpha$ for a given weight function.

In general, the distribution of recovered $(\hat{u},\hat{v})$ will be elliptical, with the shape of the ellipse described by the covariance matrix given in equation (\ref{eq:R}). In the special case that $\sigma_{u}=\sigma_{v}$ and $\sigma_{uv}=0$, the distribution will be isotropic; hence, estimates of $\hat{\alpha}$ will be unbiased. For this to be the case, when assuming $\sigma_{\mathrm{PSF}}=0$, we require
\begin{align}\label{eq:reqs}
\sum_{ij}U_{\mathrm{PSF}}^2\left(\bm{\theta}_{ij}\right)=&\sum_{ij}V_{\mathrm{PSF}}^2\left(\bm{\theta}_{ij}\right)\nonumber\\
\sum_{ij}U_{\mathrm{PSF}}\left(\bm{\theta}_{ij}\right)=&\sum_{ij}V_{\mathrm{PSF}}\left(\bm{\theta}_{ij}\right)=0,
\end{align}
as can be seen in equation (\ref{eq:cov_comp}). These requirements are satisfied if $W$ and $I_{\mathrm{PSF}}$ are both isotropic. The weight function can be chosen to be isotropic; however, the shape of the PSF is not under our control. If the PSF happens to align with the source galaxy, the estimated position angle will again be unbiased. In general, the covariance matrix will imply an elliptical distribution of $(\hat{u},\hat{v})$, and some residual bias will remain, but this bias is small enough to remain undetected in the simulations discussed in Section \ref{sec:meas}.

\label{lastpage}

\end{document}